# Instantly Decodable versus Random Linear Network Coding: A Comparative Framework for Throughput and Decoding Delay Performance

Parastoo Sadeghi, Mingchao Yu

Research School of Engineering, Australian National University, Canberra, ACT, 0200, Australia

{parastoo.sadeghi,ming.yu}@anu.edu.au

## Abstract

This paper studies the tension between throughput and decoding delay performance of two widely-used network coding schemes: random linear network coding (RLNC) and instantly decodable network coding (IDNC). A single-hop broadcasting system model is considered that aims to deliver a block of packets to all receivers in the presence of packet erasures. For a fair and analytically tractable comparison between the two coding schemes, the transmission comprises two phases: a systematic transmission phase and a network coded transmission phase which is further divided into rounds. After the systematic transmission phase and given the same packet reception state, three quantitative metrics are proposed and derived in each scheme: 1) the absolute minimum number of transmissions in the first coded transmission round (assuming no erasures), 2) probability distribution of extra coded transmissions in a subsequent round (due to erasures), and 3) average packet decoding delay. This comparative study enables application-aware adaptive selection between IDNC and RLNC after systematic transmission phase.

One contribution of this paper is to provide a deep and systematic understanding of the IDNC scheme, to propose the notion of packet diversity and an optimal IDNC encoding scheme for minimizing metric 1. This is generally NP-hard, but nevertheless required for characterizing and deriving all the three metrics. Analytical and numerical results show that there is no clear winner between RLNC and IDNC if one is concerned with both throughput and decoding delay performance. IDNC is more preferable than RLNC when the number of receivers is smaller than packet block size, and the case reverses when the number of receivers is much greater than the packet block size. In the middle regime, the choice can depend on the application and a specific instance of the problem.

## Index Terms

Packet broadcast, network coding, minimum clique cover, throughput, delay

# I. INTRODUCTION

Before the advent of network coding [1], different information flows incoming to a node in a data network were forwarded separately to other nodes in the network. By moving away from this traditional routing/scheduling approach and introducing the concept of information mixing, it was shown that network coding can achieve the information capacity of wired networks with error-free links under a single-source multicast scenario [1], [2]. Network coding can enhance the throughput performance of many other wired and wireless data networks in more general multicast or non-multicast settings and with error-free or error-prone communication links [2]–[6]. However, any statement about throughput optimality of network coding should be treated with care from a practical perspective. Roughly speaking, a higher data throughput using network coding does not necessarily translate into faster delivery of information to the application layer because the mixed information flows need to be "untangled" or network decoded first [7], [8].

Understanding the tension between throughput and delay in network coded systems has been the subject of research in recent years [7]–[27], where a packet-level network coding model is particularly suitable for such studies [8]. However, there are still many open questions to be answered and we will highlight some of them in the following.

Random linear network coding (RLNC) [28] has been one of the most successful and practical network coding schemes proposed in the literature. The encoding is very easy where a block or generation of $K_T$ data packets are combined using random coefficients from a finite field. In a *rateless* wireless broadcast scenario with packet erasures, RLNC also requires minimum amount of feedback from the receivers, as they only need to acknowledge successful reception of $K_T$ linearly independent coded packets [8]. It has been shown in [8] that rateless RLNC is throughput optimal[1] and can significantly outperform traditional scheduling approaches in terms of block completion time and hence, bandwidth or throughput efficiency. Nevertheless the fundamental tradeoff between throughput and delay still exists. A larger block size can enhance throughput and the broadcast network capacity can be arbitrarily closely approached [20]. But this comes at the cost of larger delays since packet delivery to the application layer cannot occur until block-wise network decoding is complete [7], [8], [20].[2]

This has motivated researchers to look for other network coding schemes to improve packet decoding or delivery delay. However, improved delay often comes at the cost of more frequent feedback and more complex decision making for packet encoding. For example, [14], [15] proposed network coding schemes for erasure-prone wireless broadcast, which maintain throughput optimality and use receivers' *online* feedback about every packet reception to dynamically select appropriate encoded packets for the next transmission. It was conjectured (and shown through numerical simulations) that the delay asymptotically follows that of a single-receiver case when the transmission rate approaches erasure link capacity. However, general non-asymptotic characterization of the delay performance of [14], [15] has proven to be challenging [12]. Only recently analytical results were presented on one of the three conditions for decoding [26], which were then used in [27] to propose a dynamic rate adaptation scheme. It was shown that by occasional carefully-balanced sacrifices in throughput, an overall better throughput-delay curve compared to [15] can be achieved.

---

[1] For large field sizes, we can assume that every encoded packet is linearly independent of all previously received $k < K_T$ packets [28].

[2] For large field sizes, we can ignore the probability of intermediate decoding in RLNC.

Another class of online network coding schemes which attempt to strike a balance between throughput and delay is instantly decodable network coding (IDNC) for erasure-prone wireless broadcast [9], [11], [12], [17]–[19], [22]. In a general case of IDNC [18], [19] and for a given receiver, coded packets either 1) contain one new data packet and hence are instantly decodable, 2) do not contain any new data packets and hence are non-innovative, or 3) contain more than one new data packet and hence, are deemed unusable.

One advantage of IDNC is simple XOR-based encoding and decoding. Another important advantage is that instantly decoded packets can be potentially delivered to the application layer much faster than the RLNC scheme and those discussed in [14], [15], especially if packets are usable regardless of their order [11], [17]. However, similar to [14], [15] IDNC requires full use of feedback for making online network coding decisions. In addition, IDNC scheme is generally not throughput optimal (there exist situations where a subset of receivers receive either non-innovative or unusable packets). Any attempt to provably minimize either IDNC block completion or decoding delay, even in the absence of packet erasures, results in NP-hard problems [11], [17]–[19], [22], [29]. Consequently, heuristic or stochastic approaches, corroborated with simulation, are often used in the literature to enhance and verify IDNC throughput or delay performance [9], [11], [12], [17]–[19], [22], [29]. It is fair to say that theoretical interaction between throughput and delay of IDNC is not very well understood, mainly due to the complicated nature of IDNC decision making process, especially with packet erasures.

## II. Problem Formulation and Summary of Contributions

In summary, RLNC is throughput optimal, but can incur large block decoding delays. Whereas, IDNC is not necessarily throughput optimal, but its immediate decoding property can be more beneficial for intermediate packet delivery to the application layer. The main open question is then how the throughput and delay performance of these two schemes compare. In this paper, we aim to provide a unified theoretical framework for quantitative comparison between throughput and delay performance of RLNC and IDNC schemes. To this end and to enable a meaningful comparison, we consider a transmission scenario which is briefly discussed and justified here. More detailed system model and notations will be provided in Section III.

1) There are $K_T$ packets to be broadcast to a set of $N$ wireless receivers where each receiver link is subject to independent packet erasures. Transmission is rateless, meaning that no new packet is sent until all receivers have received all $K_T$ packets.

2) In the first phase of transmission, $K_T$ uncoded packets are broadcast, which is known as the *systematic phase of network coding* [16], [23]. In order to perform IDNC in subsequent transmission phases, systematic transmission is essential in the first phase [17], [18]. This is, however, a slight variation of traditional RLNC scheme. Nevertheless, it does not result in loss of throughput optimality in RLNC and has some additional advantages [16], [23]. For example, the receivers can pass some of the uncoded received packets in a meaningful way to the application layer before the whole block of $K_T$ packets is successfully received.

3) After the systematic transmission phase, each receiver provides feedback to the sender about the packets it has successfully received or lost. Assuming that there still exist some missing packets across receivers and given the feedback from each receiver, the following key metrics about the throughput and decoding delay performance of RLNC and IDNC are determined at the sender.

   a) Metric 1: The *absolute minimum number of coded transmissions* that is needed to satisfy the demands of all receivers in RLNC and IDNC schemes in the best case scenario where there is no future packet erasures. This is denoted by $U_{\text{RLNC}}$ and $U_{\text{IDNC}}$, respectively. While $U_{\text{RLNC}}$ is simply equal to the maximum number of lost packets across all receivers, finding $U_{\text{IDNC}}$ is far from trivial. At this stage, we can only assert that $U_{\text{IDNC}} \geqslant U_{\text{RLNC}}$ due to the fact that IDNC is not necessarily throughput optimal. However, precise answer to this question cannot be determined analytically and is even hard to solve algorithmically for large $K_T$, since it requires solving an NP-hard problem [29].

b) Metric 2: The probability that there is still a need for more transmissions due to packet erasures. More specifically, the probabilities of still needing $0 \leqslant V_{\text{RLNC}} \leqslant U_{\text{RLNC}}$ and $0 \leqslant V_{\text{IDNC}} \leqslant U_{\text{IDNC}}$ extra transmissions in each scheme (after the minimum of $U_{\text{RLNC}}$ or $U_{\text{IDNC}}$ coded packets are already transmitted by the sender) are analytically computed.

c) Metric 3: The decoding performance of RLNC and IDNC schemes. More specifically, at the $u$-th coded transmission after the systematic transmission phase ($u \leqslant U_{\text{RLNC}}$ or $u \leqslant U_{\text{IDNC}}$), the probability of decoding packets at each receiver is analytically derived, from which the average packet decoding delay across all the receivers is computed.

4) Based on the derived metrics, either RLNC or IDNC scheme is adopted by the sender for subsequent coded packet transmissions. Packets are RLNC or IDNC encoded and transmitted until all $K_T$ packets are successfully received by all the receivers.

A few notes are in order about the use of feedback and predictive performance metrics in our problem formulation. First, feedback is required at the end of systematic transmission phase. This is not part of traditional RLNC. However, it enables quantitative comparison of RLNC and IDNC schemes fairly (given the same packet reception/loss status at the receivers) and choosing either scheme for the subsequent coded transmission phase. Second, metrics 2 and 3 are computed assuming that no intermediate feedback is available at the sender before the minimum of $U_{\text{RLNC}}$ or $U_{\text{IDNC}}$ coded packets are transmitted. In addition to being practically attractive for IDNC, this less frequent *semi-online* feedback enables predictive analytical comparison of RLNC and IDNC schemes. In Section IV we will prove that fundamentally, metric 1 cannot be reduced regardless of feedback frequency or packet erasures. However, we will discuss that the overall throughput of IDNC can be enhanced if *fully-online* feedback is available after each erasure-prone packet transmission. As a result, our derived performance metrics can be interpreted as lower bounds to what can be achieved in fully-online IDNC. This would actually mean a smaller throughput gap between RLNC and fully-online IDNC, as well as a better decoding delay performance gain of fully-online IDNC over RLNC.

To the best of our knowledge, such comparative framework for RLNC and IDNC schemes has not been studied before. Perhaps, the closest work to ours is [22] where the authors proposed a random graph approach to approximate metric 1 in IDNC, $U_{\text{IDNC}}$, in multicast scenarios. In contrast, we are interested in $U_{\text{IDNC}}$ given the actual deterministic IDNC graph, obtained after the first systematic transmission phase. Section IV is devoted to theoretical understanding of $U_{\text{IDNC}}$, where we provide several definitions, proofs, and bounds on the relation between $U_{\text{IDNC}}$ and the IDNC graph. We also propose an algorithm for determining $U_{\text{IDNC}}$ coded packet transmissions before semi-online feedback collection. This is needed later for computing metrics 2 and 3. While some similar (but not identical) concepts, bounds, and algorithms exist in the graph theory literature, our aim at providing a comprehensive and coherent treatment of $U_{\text{IDNC}}$ in the context of network coding is novel and useful. For example, we introduce concepts such as *packet diversity*, which can crucially affect IDNC robustness to erasures and its throughput and decoding delay performance. We also discuss the impact of feedback frequency on IDNC throughput efficiency. Finally, we highlight that both metrics 2 and 3 are new and were not studied in [22]. They will be treated in Section V and Section VI, respectively.

## III. SYSTEM MODEL AND NOTATIONS

### A. Transmission Setup

We consider a packet-based wireless broadcast scenario from one sender to $N$ receivers. Receiver $n$ is denoted by $R_n$ and the set of all receivers is $\mathcal{R} = \{R_1, \cdots, R_N\}$. There are a total of $K_T$ packets to be delivered to all receivers. Packet $k$ is denoted by $P_k$ and the set of all packets is $\mathcal{P}_{K_T} = \{P_1, \cdots, P_{K_T}\}$. Sometimes we will refer to $P_k$ as an original data packet to distinguish it from a coded packet. Time is slotted and in each time slot, one (coded or original data) packet is broadcast. The wireless channel between the sender and each receiver is modeled as a memoryless erasure link where the erasure probability for receiver $n$ is $P_{e,n}$. We further assume that the erasures across receivers are independent of each other. For simplifying the theoretical analysis later in the paper, we will impose the additional condition of equal erasure probability among links. That is, $P_{e,n} = P_e$ for $n = 1, \cdots, N$. However, the theoretical framework proposed in the paper can be generalized, with appropriate modifications, to non-homogeneous erasure links.

### B. Transmission Phases and Rounds

The minimum number of packet transmissions to satisfy the demands of all receivers is $K_T$ and there is no need to employ any type of coding for the first phase comprising $K_T$ transmissions. This is the systematic phase of network coding.

After the first $K_T$ systematic packet transmissions, each receiver provides feedback to the sender about the packets they have received or lost.[3] If a packet is successfully received by all receivers, it can be removed from any future coded packet transmission. Therefore, in the *coded transmission phase* we will deal with $K \leq K_T$ packets that are wanted by at least one receiver and the set of all such packets is $\mathcal{P}_K = \{P_1, \cdots, P_K\}$. The sender will then calculate some performance metrics and determine to use IDNC or RLNC for the subsequent phase of coded packet transmission until all receivers have received all packets. We further split the coded transmission phase into rounds. In the first round, the minimum of $U_{\text{IDNC}}$ or $U_{\text{RLNC}}$ coded packets are transmitted, which are needed in each scheme in the best case scenario of no future packet erasures. However, if due to erasures, some packets are still missing at some receivers at the end of the first round, at least one more round with $V_{\text{IDNC}} \leqslant U_{\text{IDNC}}$ or $V_{\text{RLNC}} \leqslant U_{\text{RLNC}}$ coded transmissions will be required.

*C. Receiver Demands and State Feedback Matrix*

After the systematic transmission phase, the complete state of receivers and packets can be captured by an $N \times K$ state feedback matrix (SFM) $\boldsymbol{A}$ (also known as receiver-packet incidence matrix [17]), where the element at row $n$ and column $k$ is denoted by $a_{n,k}$ and

$$a_{n,k} = \begin{cases} 1 & \text{if } R_n \text{ wants } P_k, \\ 0 & \text{otherwise.} \end{cases} \quad (1)$$

Based on the SFM, we can define the following notions of *Wants* set for each receiver and *Targeted* receivers for each packet.

**Definition 1.** *The Wants set of receiver $R_n$, denoted by $\mathcal{W}_n$, is the subset of packets in $\mathcal{P}_K$ which are lost due to erasures during systematic network coding phase and are missing at $R_n$. That is, $\mathcal{W}_n = \{P_k : a_{n,k} = 1\}$. The size of $\mathcal{W}_n$ is denoted by $W_n$.*

**Definition 2.** *The Target set of a packet $P_k$, denoted by $\mathcal{T}_k$, is the subset of receivers in $\mathcal{R}$ who want packet $P_k$. That is, $\mathcal{T}_k = \{R_n : a_{n,k} = 1\}$. The size of $\mathcal{T}_k$ is denoted by $T_k$.*

The size of the smallest and largest Wants sets among all receivers are denoted by $W_{\min}$ and $W_{\max}$, respectively, which will be key parameters in determining the system performance.

---
[3] We assume that there exists an error-free feedback link between each receiver to the sender that can be used with appropriate frequency.

## D. Coded Transmissions Using RLNC

In the case of RLNC, all the packets in $\mathcal{P}_K$ (that by definition are wanted by at least one receiver) are linearly combined with each other at the sender using randomly chosen coefficients from a finite field $\mathbb{F}_q$ of size $q$ and then broadcast to all receivers. The $r$-th coded packet is

$$X_r = \sum_{k=1}^{K} \alpha_{k,r} P_k \qquad (2)$$

where $\alpha_{k,r} \in \mathbb{F}_q$ and the coding coefficient vector for the $r$-th coded packet is $\boldsymbol{\alpha}_r = (\alpha_{1,r}, \cdots, \alpha_{K,r})^T$ and the superscript $T$ denotes vector transpose. Upon successful reception, these linearly coded packets along with their coding coefficient vectors are stored by each receiver.

**Definition 3.** *The knowledge space of each receiver after $r$ successful coded receptions is defined as the linear space spanned by the coding coefficient vectors $\boldsymbol{\alpha}_1$ to $\boldsymbol{\alpha}_r$.*

If the field size $q$ is sufficiently large, we can comfortably assume that each successful coded packet reception will be *innovative* to a receiver who has not decoded all packets yet and hence, will increase the dimension of its knowledge space by one [28]. The minimum number of coded packet transmissions for RLNC, $U_{\text{RLNC}}$, is equal to the size of the largest Wants set across all receivers, $W_{\max}$.

## E. Coded Transmissions Using IDNC

Determining suitable packets to code together that satisfy the IDNC constraints is much more complicated than RLNC. In this subsection, we first present some basic definitions related to IDNC. Then we will briefly discuss existing models in the literature to deal with the IDNC problem. Later in Section IV, we will provide a deeper understanding of the IDNC and its minimum number of coded transmissions $U_{\text{IDNC}}$, which is one of the contributions of this paper.

**Definition 4.** *A coded packet is instantly decodable for receiver $R_n$ if it contains only one original data packet that $R_n$ has not decoded yet.*

**Definition 5.** *A coded packet is non-instantly decodable for receiver $R_n$ if it includes a linear combination of two or more original data packets that $R_n$ has not decoded yet.*

**Definition 6.** *A coded packet is non-innovative for $R_n$ if it only contains original data packets that $R_n$ has already decoded. Otherwise, the coded packet is innovative.*

It suffices to use the binary field $\mathbb{F}_2$ for IDNC coded packet transmissions. More precisely, the $r$-th transmitted coded packet in the IDNC scheme is

$$Y_r = \sum_{k=1}^{K} \beta_{k,r} P_k \qquad (3)$$

where $\beta_{k,r} \in \mathbb{F}_2$ and the coding coefficients vector is $\boldsymbol{\beta}_r = (\beta_{1,r}, \cdots, \beta_{K,r})^T$ with $\boldsymbol{\beta}_r \in \{0,1\}^K$. The IDNC problem is then to determine the "best" coding coefficients vector $\boldsymbol{\beta}_r = (\beta_{1,r}, \cdots, \beta_{K,r})^T$ for each time slot $r$. This has been recognized as a highly non-trivial and computationally complex problem [11], [17]–[19], [22], [29] and there have been various approaches in the literature for solving this problem, which we now briefly discuss.

*F. S-IDNC versus G-IDNC*

One main model to capture IDNC constraints and to determine coded packets is strict IDNC or S-IDNC [11], [17], [29]. Imposing the S-IDNC constraints means that for a given receiver $R_n$, $Y_r$ can at most contain one packet from the Wants set of that receiver. Mathematically, either $\sum_{k:a_{n,k}=1} \beta_{k,r} = 1$ resulting in decoding one new packet at $R_n$ or $\sum_{k:a_{n,k}=1} \beta_{k,r} = 0$ which renders $Y_r$ non-innovative for $R_n$. The S-IDNC constraints for all receivers can be concisely written using the SFM $\boldsymbol{A}$ as $\boldsymbol{A}\boldsymbol{\beta}_r \leqslant \boldsymbol{1}_N$, where $\boldsymbol{1}_N$ is an all-one $N \times 1$ vector and the inequality means that every entry in $\boldsymbol{A}\boldsymbol{\beta}_r$ is smaller or equal to the corresponding entry in $\boldsymbol{1}_N$. As has been mentioned in [29], the S-IDNC constraints imposed by the SFM, $\boldsymbol{A}$, can also be represented using an undirected graph with $K$ vertices corresponding to $K$ wanted packets. In Section IV, we will examine the graphical representation of S-IDNC in detail and provide new insights into the S-IDNC problem.

S-IDNC is a strong type of IDNC. Intuitively, its main limitation is that some coding opportunities benefiting a large number of receivers may be lost just because a small number of receivers cannot instantly decode those coded packets. This can potentially reduce the throughput. In contrast, G-IDNC proposed in [18], [19] relaxes the problem to a certain extent with the aim to improve throughout and decoding delay (compared to [17]), while still benefiting from simple XOR-based decoding at the receivers. The sender may send packets that are instantly decodable only for an appropriately chosen subset of receivers. If a receiver receives a coded packet which is non-instantly decodable to it, it discards that packet. In other words, the sender is not restricted to send fully IDNC packets for every receiver, but the receivers adhere to IDNC decoding principle.

Obtaining the corresponding G-IDNC graph from the SFM $\boldsymbol{A}$ is detailed in [18], [19], [22], where the authors use packet (fully-online) feedback to update the graph after each transmission and select a *maximal clique*[4] through a heuristic maximum weighted vertex search. However, the size of G-IDNC graph is in general $\mathcal{O}(NK)$ as opposed to the size of S-IDNC graph, which is $K$. Hence, finding the minimum number of coded transmissions for G-IDNC (even with no erasures) is much more computationally expensive than S-IDNC. This is why a random graph approach has been adopted in [22] to adaptively select between RLNC and G-IDNC in multicast scenarios.

In the rest of this paper, our aim will be to better understand and characterize the S-IDNC problem and compare it with RLNC from a theoretical viewpoint. We will simply refer to S-IDNC as IDNC. Based on the reported results in [18], [19], we expect that the throughput performance of G-IDNC is superior to S-IDNC. Hence, the study of this paper can serve as worst-case throughput gap between G-IDNC and RLNC, as well as worst-case decoding delay performance gain of G-IDNC over RLNC.

## IV. Understanding IDNC

As has been mentioned in the last section, IDNC constraints of an SFM $\boldsymbol{A}$ can be represented by an undirected graph $\mathcal{G}(\mathcal{V}, \mathcal{E})$ with $K$ vertices. Each vertex $v_i \in \mathcal{V}$ represents a wanted original packet $P_i$ and two vertices $v_i$ and $v_j$ are connected by an edge $e_{i,j} \in \mathcal{E}$ if $P_i$ and $P_j$ are not jointly wanted by any receiver [29]. This graph model, however, has only been employed to heuristically find IDNC encoding solutions in the literature [29]. In this section, we will revisit this graph model and its equivalent matrix/set representation. We provide several definitions that facilitate our analysis of the IDNC performance in this and the next two sections. We then prove the relation between the minimum number of coded transmissions, $U_{\text{IDNC}}$, and the IDNC graph and discuss the effect of feedback frequency on IDNC throughput. We also present two simple bounds on $U_{\text{IDNC}}$. Finally, we propose an IDNC encoding algorithm which is optimum in terms of $U_{\text{IDNC}}$. Although some similar concepts and results exist in the graph theory literature, their compilation, presentation and more importantly interpretation in the IDNC context is new, to the best of our knowledge. We will highlight the similarities, differences, and new results as appropriate.

---

[4]In an undirected graph, all vertices in a clique are connected to each other with an edge and a clique is maximal if no other vertex can be added to it to form a larger clique.

*A. IDNC Modeling*

In this subsection, we construct the matrix/set model of IDNC and demonstrate its relationship with the graph model. The construction of the matrix/set model is based on the concept of conflicting packets, which is defined below.

**Definition 7.** *We say that two packets $P_i$ and $P_j$ conflict with each other if there exists at least one receiver who wants both packets. That is, $P_i$ and $P_j$ conflict if both belong to the Wants set, $\mathcal{W}_n$, of at least one receiver such as $R_n$. Mathematically, we can denote a conflict between $P_i$ and $P_j$ by $P_i \oslash P_j$, where $P_i \oslash P_j \Leftrightarrow \exists n : \{P_i, P_j\} \subset \mathcal{W}_n$.*

It is clear that based on the requirement of IDNC, two conflicting packets $P_i$ and $P_j$ cannot be encoded together. The equivalent of such conflict in the graph model is the absence of an edge between the two vertices [29], [30].

**Definition 8.** *We say that two packets $P_i$ and $P_j$ do not conflict with each other if there is no receiver who wants both packets. Mathematically, we can denote a lack of conflict between $P_i$ and $P_j$ by $P_i \overline{\oslash} P_j$, where $P_i \overline{\oslash} P_j \Leftrightarrow \nexists n : \{P_i, P_j\} \subset \mathcal{W}_n$.*

The equivalent of two non-conflicting packets in the graph model is an edge between the two vertices [29], [30].

Based on Definition 7 and 8, we can define a triangular conflict matrix $\boldsymbol{C}$ of size $K(K-1)/2$ that can fully describe the conflict state of packets based on the SFM $\boldsymbol{A}$ and is given below.

**Definition 9.** *A fully-square conflict matrix of size $K^2$ is a binary-valued matrix with element at row $i$ and column $j$ denoted by $c_{i,j}$ corresponding to the conflict state of packets $P_i$ and $P_j$. In particular, $c_{i,j} = 1$ if $P_i \oslash P_j$ and $c_{i,j} = 0$ if $P_i \overline{\oslash} P_j$.*

*Due to the symmetry of conflict between packets and noting that $c_{i,i} = 0$, $\forall P_i$, we can reduce the fully-square conflict matrix to a triangular matrix $\boldsymbol{C}$ of size $K(K-1)/2$. From now on, by conflict matrix, we mean the reduced triangular matrix $\boldsymbol{C}$.*

**Example 1.** *An example of a conflict matrix is given below.*

$$C = \begin{array}{c|ccccc|c} & P_2 & P_3 & P_4 & P_5 & P_6 & \\ \hline & 0 & 0 & 0 & 1 & 1 & P_1 \\ & & 0 & 1 & 0 & 1 & P_2 \\ & & & 1 & 1 & 0 & P_3 \\ & & & & 1 & 1 & P_4 \\ & & & & & 1 & P_5 \end{array}$$

Note that in dealing with the conflict matrix, we are not concerned with the receivers who may need a certain packet. In fact, it is not difficult to show that two or more SFMs can have the same conflict matrix. Unless otherwise stated, it suffices to deal with conflict matrix $C$, instead of SFM $A$.

We now define the key concept of *maximal encoding sets* that are allowed for IDNC transmission.

**Definition 10.** *A maximal encoding set, $\mathcal{M}$, is a collection of original packets which simultaneously hold the following two properties: 1) their coded transmission do not violate the instant decodability constraint for any receiver $R_n$. That is, $\sum_{P_i \in \mathcal{M}} P_i$ is either instantly decodable or non-innovative for every receiver $R_n \in \mathcal{R}$, where the sum is over $\mathbb{F}_2$; 2) addition of any other packet from $\mathcal{P}_K \setminus \mathcal{M}$ to $\mathcal{M}$ will result in the violation of instantly decodable property of $\sum_{P_i \in \mathcal{M}} P_i$ for at least one receiver such as $R_n \in \mathcal{R}$.*

Based on the definition of conflicting and non-conflicting packets, we can say that firstly none of the packets in a maximal encoding set $\mathcal{M}$ should conflict with each other. That is, $\forall i, j : \{P_i, P_j\} \subset \mathcal{M} \Rightarrow P_i \overline{\oslash} P_j$. Furthermore, addition of any packet such as $P_k$ from $\mathcal{P}_K \setminus \mathcal{M}$ to $\mathcal{M}$ will result in conflict(s) with its existing member(s). That is, $\forall P_k \in \mathcal{P}_K \setminus \mathcal{M}, \exists P_j \in \mathcal{M} : P_k \oslash P_j$. The equivalent of maximal encoding sets in the IDNC graph model is known as maximal cliques [30].

In Example 1, it can be easily verified that $\{P_1, P_2, P_3\}$, $\{P_1, P_4\}$, $\{P_2, P_5\}$ and $\{P_3, P_6\}$ are all maximal encoding sets.

For reasons that become clear at the end of this subsection, we ensure that in each IDNC coded packet transmission, the sender will encode *all and not a subset of* the original data packets in a maximal encoding set. To satisfy the demand of all the receivers, the sender has to transmit encoded packets from an appropriately chosen collection of maximal encoding sets. To achieve this, each original data packet should appear at least once in this collection. This condition can be formally represented by the *diversity constraint*, where:

**Definition 11.** *The diversity of a packet $P_i$ within a collection of maximal encoding sets is denoted by $d_i$ and is the number of maximal encoding sets in which it appears.*

**Definition 12.** *A collection of maximal encoding sets satisfies the diversity constraint iff every original packet has a diversity of at least one within this collection.*

For example, given three maximal encoding sets: $\mathcal{M}_1 = \{P_1, P_2, P_3\}$, $\mathcal{M}_2 = \{P_1, P_4\}$, $\mathcal{M}_3 = \{P_2, P_5\}$, we have $d_1 = d_2 = 2$, $d_3 = d_4 = d_5 = 1$.

Given all the maximal encoding sets of a conflict matrix $C$, there exists at least one collection which satisfies the diversity constraint (in the extreme case all the maximal encoding sets include all the original packets). The size of the collection is the number of maximal encoding sets in it. We then define the *minimal collection* and its size as follows:

**Definition 13.** *A collection of maximal encoding sets is minimal if there does not exist any collection which satisfies the diversity constraint with a smaller size. The size of the minimal collection is called the minimum collection size. This number, as we will prove in the next subsection, is exactly the absolute minimum number of coded transmissions, $U_{\text{IDNC}}$.*

We denote a minimal collection by $\mathcal{S} = \{\mathcal{M}_1, \cdots, \mathcal{M}_{U_{\text{IDNC}}}\}$. A problem in the graph theory which is *similar* to finding a minimal collection of maximal encoding sets is the "minimum clique cover" problem [29], [30]. These two problems differ somewhat because cliques do not overlap in the minimum clique cover problem. That is, the cliques are not necessarily maximal and each vertex appears in only one clique. This would be equivalent to choosing a minimal collection of encoding sets in our IDNC solution where all packets have a diversity equal to 1. This would not change $U_{\text{IDNC}}$. However, it can have a serious adverse impact on IDNC robustness to erasures, which in turn degrades the IDNC overall throughput and decoding delay performance. Consequently, it is desirable to choose a minimal collection of *maximal* encoding sets that, while satisfying minimum $U_{\text{IDNC}}$, provides as much *packet diversity* as possible. This will be further clarified in Section IV-D, Section V and Section VI.

*B. The Equivalence between $U_{\text{IDNC}}$ and the Minimum Collection Size*

In this subsection, we prove that the absolute minimum number of coded transmissions, $U_{\text{IDNC}}$, equals the minimum collection size of the IDNC conflict matrix. We will also comment on the role of feedback frequency on the number of coded transmissions in IDNC.

For a given conflict matrix $\boldsymbol{C}$, proving the equivalence between the absolute minimum number of coded transmissions and the minimum collection size is equivalent to proving the following theorem:

**Theorem 1.** *Upon successful reception of a maximal encoding set $\mathcal{M}$ of $\boldsymbol{C}$ by all the target receivers of $\mathcal{M}$, the minimum collection size of the updated conflict matrix $\boldsymbol{C}'$ is at least $U_{\mathrm{IDNC}} - 1$.*

This theorem holds if the following two theorems hold:

**Theorem 2.** *The minimum collection size of a conflict matrix $\boldsymbol{C}$ with a graph model $\mathcal{G}$ equals the chromatic number of the complementary graph of $\mathcal{G}$, denoted by $\overline{\mathcal{G}}$, whose vertex connectivity is opposite to $\mathcal{G}$.*

**Theorem 3.** *Suppose $\mathcal{M}$ is a maximal clique of $\mathcal{G}$ and the chromatic number of $\overline{\mathcal{G}}$ is $U_{\mathrm{IDNC}}$. By removing the vertices in $\mathcal{M}$ from $\mathcal{G}$ we obtain an updated graph $\mathcal{G}'$. The chromatic number of $\overline{\mathcal{G}'}$ is at least $U_{\mathrm{IDNC}} - 1$.*

The $\mathcal{G}'$ here is indeed the graph model of the updated $\boldsymbol{C}'$ because if $\mathcal{M}$ has been successfully received by all its targeted receivers, original packets in $\mathcal{M}$ are no longer wanted. That is, the corresponding vertices can be removed from $\mathcal{G}$. The proofs of Theorem 2 and 3 are provided in Appendices A-B.

**Remark 1.** *Greedy/heuristic algorithms cannot guarantee $U_{\mathrm{IDNC}}$. In the proof of Theorem 3 we mentioned that if a maximal encoding set $\mathcal{M}$, which does not belong to any minimal collection of the conflict matrix $\boldsymbol{C}$ is successfully received by its target receivers, the chromatic number of the updated $\overline{\mathcal{G}'}$ is still $U_{\mathrm{IDNC}}$. This means that the updated matrix $\boldsymbol{C}'$ still needs $U_{\mathrm{IDNC}}$ transmissions. For example, consider the maximal encoding sets in Example 1. A suboptimal IDNC algorithm might choose $\{P_1, P_2, P_3\}$, which does not belong to the only minimal collection $\mathcal{S} = \{\{P_1, P_4\}, \{P_2, P_5\}, \{P_3, P_6\}\}$. Even if this set is successfully received, three further transmissions are needed for $P_4, P_5$ and $P_6$. In total there will be 4 transmissions, greater than $U_{\mathrm{IDNC}} = 3$.*

Assume that the sender chooses IDNC over RLNC scheme and finds an appropriate $\mathcal{S}$. Clearly, the order of transmission of maximal encoding sets within $\mathcal{S}$ does not alter $U_{\mathrm{IDNC}}$, but it can affect the decoding delay performance of IDNC and will be discussed in Example 3 and Section VI. In what follows, we assume that the maximal encoding sets within $\mathcal{S}$ are appropriately ordered. Then all packets in set $\mathcal{M}_1$ are added together in $\mathbb{F}_2$ and transmitted in the first coded transmission. The process is repeated until all packets in $\mathcal{M}_{U_{\mathrm{IDNC}}}$ are encoded and transmitted.

According to the above theorems, collecting fully online feedback during $U_{\text{IDNC}}$ transmissions cannot reduce the total number of coded transmissions below $U_{\text{IDNC}}$, even in the best case scenario of erasure-free packet reception. Hence, as a variation of existing IDNC schemes in the literature, one can reduce feedback frequency to *semi-online* feedback, where the SFM $\boldsymbol{A}$ is updated in rounds. For example, after $U_{\text{IDNC}}$ encoded packets from a selected minimal collection $\mathcal{S}$ have been transmitted and so on. In fact, the systematic phase of IDNC is a special case in this scheme: here the SFM $\boldsymbol{A}$ is a $K_T \times K_T$ all-one matrix, which requires that all the $K_T$ original packets should be transmitted separately in $U_{\text{IDNC}} = K_T$ transmissions before the sender collects feedback.

In addition to making the throughput and delay analysis of IDNC tractable, a lower feedback frequency can be advantageous in practical implementation of IDNC where the use of reverse link is costly and involves transmission of some control overheads. Another practical attraction is that it also avoids solving the computationally expensive IDNC encoding problem in every time slot. However, this comes at the potential cost of overall throughput inefficiency in semi-online IDNC, as we explain below.

Now imagine an IDNC scheme in the presence of erasures. In the semi-online feedback case, the sender sends the entire maximal encoding sets in $\mathcal{S}$ before collecting feedback and solving the IDNC encoding problem again. Let us assume that in total, $X$ IDNC transmission rounds are required with a total of $U_{\text{IDNC}}^{\text{semi-online}} = \sum_{x=1}^{X} U_{\text{IDNC}}^{x}$ coded transmissions. In contrast, an *optimal* fully online IDNC scheme works as follows. It updates the SFM $\boldsymbol{A}$ after each coded transmission and constructs the corresponding conflict matrix $\boldsymbol{C}$. It then finds $U_{\text{IDNC}}$ and a minimal collection $\mathcal{S}$ for $\boldsymbol{C}$. However, it only sends the encoded packets in $\mathcal{M}_1$ from $\mathcal{S}$ before collecting another feedback from the receivers and solving the IDNC encoding problem again. The sender repeats the process until all packets are received by all receivers. Let us denote the total number of coded transmissions in this case by $U_{\text{IDNC}}^{\text{online}}$. One can find examples where $U_{\text{IDNC}}^{\text{online}} < U_{\text{IDNC}}^{\text{semi-online}}$, where one such example is provided in Appendix C. However, characterizing performance differences between fully online and semi-online IDNC is still an open question and beyond the scope of this paper. Intuitively, we expect the difference to be small when packet erasure probability is low. In any case, our throughput and delay analysis of IDNC scheme based on semi-online feedback serves as a worst-case scenario for an optimal fully-online IDNC with packet erasures.

## C. Two bounds on $U_{\text{IDNC}}$

Here, we present an upper bound and a lower bound on $U_{\text{IDNC}}$, which only depend on the number of zeros in the conflict matrix $C$. These bounds are important references for practical/heuristic IDNC encoding algorithm design: any algorithm offering $U_{\text{IDNC}}$ above the upper bound or below the lower bound is throughput inefficient or non-instantly decodable, respectively.

For any $N \times K$ SFM $A$, its conflict matrix $C$ fully represents the corresponding IDNC encoding problem. Thus $U_{\text{IDNC}}$ only depends on the number (denoted by $M_0$) and positions of zeros in $C$. Although determining $U_{\text{IDNC}}$ for a $C$ with an arbitrary positioning of $M_0$ zeros is NP-hard, by deliberately positioning these zeros we are able to find out the minimum and maximum possible $U_{\text{IDNC}}$. Both the upper- and lower-bound depend only on $K$ and $M_0$ and are denoted by $U_{\text{IDNC}}(K, M_0)^+$ and $U_{\text{IDNC}}(K, M_0)^-$, respectively.

- $U_{\text{IDNC}}(K, M_0)^+$

  The general intuition here is *trying our best to waste the encoding opportunities brought by the $M_0$ zeros*. We first note that for any given packet, there are $K - 1$ entries in $C$ about the conflict of that packet with all other packets. When $M_0 = 0$, there is no encoding opportunities, so $U_{\text{IDNC}}(K, M_0 = 0)^+ = K$. When $M_0$ belongs to the closed interval $[1, K - 1]$, we can assign all zeros to the entries about the same packet, say $P_1$, to increase its encoding opportunities. But $U_{\text{IDNC}}$ remains $K - 1$ because $P_2, \cdots, P_K$ have to be transmitted separately. After $K - 1$ zeros have been exhausted, there are $K - 2$ entries in $C$ about every packet other than $P_1$. Thus when $M_0$ belongs to the closed interval $[K, (K-1) + (K-2)]$, we can assign these extra $K - 2$ zeros to the entries about the same packet, say $P_2$, and $U_{\text{IDNC}}$ remains $K - 2$ because $P_3, \cdots, P_K$ have to be transmitted separately. This iterative process indicates that $U_{\text{IDNC}}(K, M_0)^+$ decreases in a staircase way with $M_0$. The relationship can be expressed as:

  $$U_{\text{IDNC}}(K, M_0)^+ = \begin{cases} K, & M_0 = 0 \\ K - 1, & M_0 \in [1, K - 1] \\ K - 2, & M_0 \in [K, 2K - 3] \\ \vdots & \vdots \\ 1, & M_0 = K(K-1)/2 \end{cases} \qquad (4)$$

  and is shown in Fig. 1. This bound can also be derived from an upper-bound on the chromatic number of the corresponding complementary graph $\overline{\mathcal{G}}$:

$$\chi(\overline{\mathcal{G}}) \leqslant \Delta(\overline{\mathcal{G}}) \tag{5}$$

where $\Delta(\overline{\mathcal{G}})$ is the maximum degree of $\overline{\mathcal{G}}$. Suppose every original packet $P_i$ has $\delta_i$ nonzero entries in $\boldsymbol{C}$ about it, then $\Delta(\overline{\mathcal{G}}) = \max_i(\delta_i)$.

- $U_{\text{IDNC}}(K, M_0)^-$

The intuition here is *making the best of the encoding opportunities brought by the $M_0$ zeros*. In other words, we should *use as few zeros as possible to reduce $U_{\text{IDNC}}$ by one*. When $M_0 = 0$, no packets can be encoded together. Thus $\mathcal{S} = \{\{P_1\}, \{P_2\}, \cdots, \{P_K\}\}$ and $U_{\text{IDNC}} = K$. We then reduce $U_{\text{IDNC}}$ iteratively. In each iteration, $U_{\text{IDNC}}$ can be reduced by 1, i.e., the size of $\mathcal{S}$ can be reduced by 1, if we can merge two maximal encoding sets in $\mathcal{S}$ together. In order to use as few zeros as possible, we always pick two sets with the smallest sizes (say $m$ and $n$) to merge, which requires $mn$ zeros to resolve the conflicts between the packets in them. Hence, in each iteration, it is impossible to reduce $U_{\text{IDNC}}$ by 1 until $mn$ new zeros are added to $\boldsymbol{C}$. This iterative process provides a lower-bound of $U_{\text{IDNC}}(K, M_0)$. Below is an example with $K = 5$:

**Example 2.**

*In the first iteration, we can reduce the size of $\mathcal{S}$ by 1 by merging $\{P_1\}$ and $\{P_2\}$ together, which requires one zero, as in Eq. (6a). In the second iteration, we can reduce the size of $\mathcal{S}$ by one by merging $\{P_3\}$ and $\{P_4\}$ together, which requires 1 zero, as in (6b). In the third iteration, $2 \times 1 = 2$ zeros are needed to merge the two smallest maximal encoding sets $\{P_1, P_2\}$ and $\{P_5\}$ together because we have to resolve the conflicts between $P_1$ and $P_5$ and between $P_2$ and $P_5$. In the last iteration, $3 \times 2 = 6$ zeros are needed to merge $\{P_1, P_2, P_5\}$ and $\{P_3, P_4\}$ together. After that, all the 10 entries in $\boldsymbol{C}$ become zeros and $U_{\text{IDNC}}$ becomes 1.*

$$\mathcal{S} = \{\{P_1\}, \{P_2\}, \{P_3\}, \{P_4\}, \{P_5\}\} \quad U_{\text{IDNC}} = 5$$

$$\mathcal{S} = \{\{P_1, P_2\}, \{P_3\}, \{P_4\}, \{P_5\}\} \quad 1 \times 1 = 1 \text{ zero needed}, U_{\text{IDNC}} = 4 \tag{6a}$$

$$\mathcal{S} = \{\{P_1, P_2\}, \{P_3, P_4\}, \{P_5\}\} \quad 1 \times 1 = 1 \text{ zero needed}, U_{\text{IDNC}} = 3 \tag{6b}$$

$$\mathcal{S} = \{\{P_1, P_2, P_5\}, \{P_3, P_4\}\} \quad 2 \times 1 = 2 \text{ zeros needed}, U_{\text{IDNC}} = 2 \tag{6c}$$

$$\mathcal{S} = \{\{P_1, P_2, P_3, P_4, P_5\}\} \quad 2 \times 3 = 6 \text{ zeros needed}, U_{\text{IDNC}} = 1 \tag{6d}$$

*Similar to the upper-bound, the lower-bound also decreases in a staircase way with $M_0$. In this example:*

$$U_{\text{IDNC}}(5, M_0)^- = \begin{cases} 5, & M_0 = 0 \\ 4, & M_0 = 1 \\ 3, & M_0 \in [2, 3] \\ 2, & M_0 \in [4, 9] \\ 1, & M_0 = 10 \end{cases} \qquad (7)$$

This relationship can be approximated by a single formula which is derived by Geller in [31], [32] using a different approach:

$$U_{\text{IDNC}}(K, M_0)^- = \left\lceil \frac{K^2}{K + 2M_0} \right\rceil \qquad (8)$$

where $\lceil m \rceil$ denotes the smallest integer greater than $m$ and one can easily verify that our bound is tighter.

Although the upper- and lower-bounds of $U_{\text{IDNC}}(K, M_0)$ can be determined, there does not exist a simple theoretical approach to find out the distribution of $U_{\text{IDNC}}(K, M_0)$ between the bounds. The best we can do is to run numerical simulations to obtain the average of $U_{\text{IDNC}}(K, M_0)$. Fig. 1 demonstrates the bounds and numerical average of $U_{\text{IDNC}}(K, M_0)$ with $K = 20$ wanted original packets. For each value of $M_0$, random permutations of $M_0$ zeros and $K(K-1)/2 - M_0$ ones are assigned to the entries of $C$. Optimum IDNC algorithm which will be proposed in the next subsection is then performed to determine the $U_{\text{IDNC}}$ for this $C$.

While the upper-bound decreases gradually with $M_0$, the lower-bound falls down very quickly at first. With as few as 30 zero entries, it is possible to reduce $U_{\text{IDNC}}$ from 20 to only 5. After that, the lower-bound decreases much more slowly. It requires another 16 zero entries to reduce the lower-bound from 5 to 4 and from 4 to 3. This number is 32 for reduction of $U_{\text{IDNC}}(K, M_0)^-$ from 3 to 2, and is as large as 96 for its reduction from 2 to 1. The largest gap between the two bounds is 13, taking place when $M_0 = [30, 37]$. The average of $U_{\text{IDNC}}(K = 20, M_0)$, without surprise, lies between the two bounds. It decreases smoothly with $M_0$ and is closer to the lower-bound than to the upper-bound.

*D. Optimum IDNC Encoding Algorithm*

An optimum IDNC encoding algorithm should provide a solution which satisfies the demand of all the receivers with $U_{\text{IDNC}}$ transmissions in the best case scenario with no packet erasures. In our model, the algorithm should

19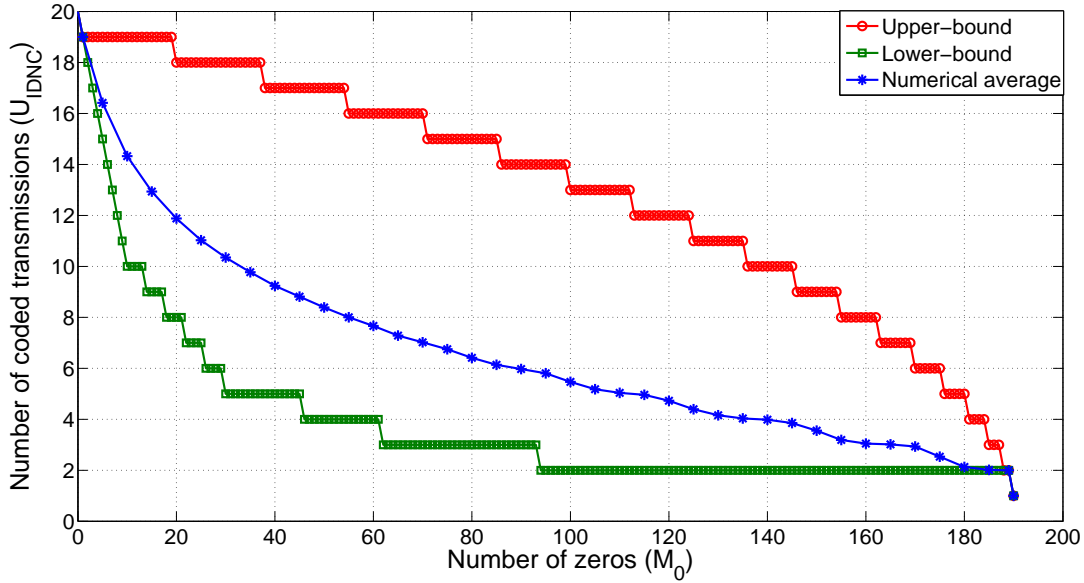

**Fig. 1:** Theoretical bounds and numerical average of $U_{\text{IDNC}}$ against $M_0$ with $K = 20$.

find minimal collections of the conflict matrix $C$, which is similar to the minimum clique cover problem in graph theory and is generally NP-hard [29], [33]. In [29], the authors proposed an optimum algorithm based on exhaustive search: all the $2^K$ possible encoding sets (whether instantly decodable or not) were generated. Then all the size-$1, 2, \cdots$ collections of these sets were examined to see if any of them satisfy both the IDNC constraints and the diversity constraint. The process ended after all the size-$U_{\text{IDNC}}$ collections of these sets were examined. However, this algorithm is highly computationally expensive even with a medium-size $\mathcal{P}_K$ [29]. In this subsection, we propose to partition this task into two stages to reduce the computational load. Further reduction can be achieved if the second stage is replaced by a heuristic algorithm. The two-stage process is as follows:

1) **Finding all the maximal cliques:** An efficient algorithm is Bron-Kerbosch algorithm [34] which finds all the maximal cliques of an undirected graph in a recursive way. We use this algorithm to obtain all the maximal encoding sets. The group of all maximal encoding sets is denoted by $\mathcal{A}$.

2) **Finding minimal collections from $\mathcal{A}$:** There is no polynomial time algorithm for this task. We propose an iterative algorithm which is much faster than exhaustive search. The first intuition here is that if a packet belongs to only one of the maximal encoding sets in $\mathcal{A}$, this set has to be transmitted. A more general intuition is that if a packet belongs to $d$ maximal encoding sets, one of these $d$ sets has to be transmitted.

**Remark 2.** *After obtaining all the maximal encoding sets in the first stage, a simple heuristic algorithm can be employed to find a collection instead of Algorithm 1. For example, in a greedy algorithm, we first choose the*

**Algorithm 1** Finding minimal collections from $\mathcal{A}$

1: Generate a group $\mathcal{B}$ of collections which contains only an empty collection $\mathcal{S}$, and generate an iteration counter $u = 1$.
2: **while** None of $\mathcal{S}$ in $\mathcal{B}$ satisfies the diversity constraint **do**
3:    **while** Not all collections in the group $\mathcal{B}$ have a size of $u$ **do**
4:       Pick a collection in $\mathcal{B}$ with size $u-1$, say $\mathcal{S} = \{\mathcal{M}_1, \cdots, \mathcal{M}_{u-1}\}$. Denote the original packets within $\mathcal{S}$ by $\mathcal{P}_{u-1} = \bigcup_{i:\mathcal{M}_i \in \mathcal{S}}\{\mathcal{M}_i\}$ and all the remaining original packets by $\overline{\mathcal{P}}_{u-1} = \mathcal{P}_K \setminus \mathcal{P}_{u-1}$. Also denote the group of maximal encoding sets of $C$ other than $\mathcal{M}_1, \cdots, \mathcal{M}_{u-1}$ by $\overline{\mathcal{S}} = \mathcal{A} \setminus \mathcal{S}$.
5:       Pick a packet, say $P$, in $\overline{\mathcal{P}}_{u-1}$ which has the smallest diversity $d$ within $\overline{\mathcal{S}}$ and find out all the maximal encoding sets comprising it, say $\mathcal{M}'_1, \cdots, \mathcal{M}'_d$;
6:       Branch $\mathcal{S}$ into $d$ new collections, $\mathcal{S}'_1, \cdots, \mathcal{S}'_d$. Then add $\mathcal{M}'_1, \cdots, \mathcal{M}'_d$ to them respectively. These collections have a size of $u$ and replace $\mathcal{S}$ in $\mathcal{B}$.
7:    **end while**
8:    $u = u + 1$
9: **end while**

largest maximal encoding set. Then for the remaining packets not belonging to this set, we look for a set which comprises most of them. This iterative algorithm is much faster than Algorithm 1, but can be suboptimal in terms of $U_{\text{IDNC}}$. There are also other heuristic algorithms (based on graph model or not) in the literature, interested readers are referred to [29].

**Remark 3.** *For a given SFM $\mathbf{A}$ we find all the minimal collections of its corresponding conflict matrix $\mathbf{C}$ through Algorithm 1. If there are more than one such collection, we can select the one having larger diversities for the packets wanted by more receivers to gain the highest system robustness to packet erasures. Our proposed selection criteria for a minimal collection $\mathcal{S}$, which is a scalar value and denoted by $\sigma$, is given by*

$$\sigma(\mathcal{S}) = \sum_{k=1}^{K} d_k T_k \qquad (9)$$

*where $d_k$ is the diversity of $P_k$ within $\mathcal{S}$ and $T_k$ is the number of targeted receivers of $P_k$. Then the transmission order of the maximal encoding sets in this collection could be adjusted to benefit more receivers first.*

In this paper, we will always use the above process to find the minimal collection of $A$ which has the largest average diversity and has its maximal encoding sets reordered based on maximum decoding benefit to receivers. This minimal encoding set is *the* minimal collection of this $A$ for transmission and used for throughput and delay analysis. Below is an example of the proposed two-stage IDNC encoding process:

**Example 3.**

$$A = \begin{array}{c|cccccc|c} & P_1 & P_2 & P_3 & P_4 & P_5 & P_6 & \\ \hline & 1 & 0 & 0 & 1 & 1 & 0 & R_1 \\ & 1 & 0 & 0 & 0 & 0 & 1 & R_2 \\ & 1 & 1 & 0 & 0 & 0 & 1 & R_3 \\ & 1 & 1 & 0 & 1 & 0 & 0 & R_4 \\ & 0 & 0 & 1 & 0 & 0 & 1 & R_5 \end{array}$$

$$C = \begin{array}{c|ccccc|c} & P_2 & P_3 & P_4 & P_5 & P_6 & \\ \hline & 1 & 0 & 1 & 1 & 1 & P_1 \\ & & 0 & 1 & 0 & 1 & P_2 \\ & & & 0 & 0 & 1 & P_3 \\ & & & & 1 & 0 & P_4 \\ & & & & & 0 & P_5 \end{array}$$

*Given the SFM, $A$ and its corresponding conflict matrix $C$, it can be easily verified that the maximal encoding sets are $\{P_1, P_3\}$, $\{P_2, P_3, P_5\}$, $\{P_3, P_4\}$, $\{P_4, P_6\}$ and $\{P_5, P_6\}$. Then the algorithm will find the minimum collection(s) as follows:*

- *In the first iteration, $P_1$ with $d_1 = 1$ is picked, thus $\mathcal{S} = \{\{P_1, P_3\}\}$. This collection does not satisfy the diversity constraint;*

- *In the second iteration, the remaining packets are $P_2, P_4, P_5$ and $P_6$. Then $P_2$ with $d_2 = 1$ is picked. Thus $\mathcal{S} = \{\{P_1, P_3\}, \{P_2, P_3, P_5\}\}$. This collection still does not satisfy the diversity constraint;*

- *In the third iteration, the remaining packets are $P_4$ and $P_6$, with $d_4 = d_6 = 2$. Then $P_4$ is picked. Thus $\mathcal{S}_1 = \{\{P_1, P_3\}, \{P_2, P_3, P_5\}, \{P_3, P_4\}\}$ and $\mathcal{S}_2 = \{\{P_1, P_3\}, \{P_2, P_3, P_5\}, \{P_4, P_6\}\}$. Because $\mathcal{S}_1$ does not satisfy the diversity constraint but $\mathcal{S}_2$ does, $\mathcal{S}_2$ is the minimal collection and $U_{\text{IDNC}} = 3$.*

- *Because there is only one minimal collection, no additional selection process is needed.*

- *We then can reorder $\mathcal{S}_2$ into $\mathcal{S} = \{\{P_1, P_3\}, \{P_4, P_6\}, \{P_2, P_3, P_5\}\}$ to benefit 5, 5, and 3 receivers in the first, second and third transmissions, respectively.*

We remark that other selection and reordering approaches can be easily adopted if higher priorities are issued to specific packets or receivers, which depends on the application.

## V. THE DISADVANTAGE OF IDNC OVER RLNC: THROUGHPUT ANALYSIS

Since rateless transmission is assumed, system throughput is inversely proportional to the total number of transmissions required to successfully broadcast the current block of $K_T$ original data packets to all the receivers. This number consists of three parts:

1) In the systematic transmission phase, $K_T$ uncoded original data packets are transmitted. This number is fixed regardless of whether we use RLNC or IDNC later. Therefore, we will not consider it in this section.

2) In the first round of coded transmission phase, a minimum of $U_{\text{IDNC}}$ and $U_{\text{RLNC}}$ transmissions are required by IDNC and RLNC, respectively. In general, we have $U_{\text{IDNC}} \geqslant U_{\text{RLNC}}$. In this section, we will investigate how the number of receivers affects the gap between $U_{\text{IDNC}}$ and $U_{\text{RLNC}}$.

3) After the first round of coded transmission, extra transmissions may be still needed due to erasures. We denote the number of extra transmissions in the second round by $V_{\text{IDNC}}$ and $V_{\text{RLNC}}$ for IDNC and RLNC, respectively. In this section, we will derive their probability distribution and verify the derivations through simulations.

By combining two coded transmission rounds (that is by studying $U_{\text{IDNC}}$ and $V_{\text{IDNC}}$ together and $U_{\text{RLNC}}$ and $V_{\text{RLNC}}$ together), we can shed light on the relative throughput performance of IDNC and RLNC.

### A. Comparing $U_{\text{IDNC}}$ and $U_{\text{RLNC}}$

Our objective here is to compare the mean of $U_{\text{IDNC}}$ and the mean of $U_{\text{RLNC}}$ under the same system parameters: total packet number $K_T$, link erasure probability $P_e$, and number of receivers $N$. By changing $N$, we are able to observe how the number of receivers affects the gap between $U_{\text{IDNC}}$ and $U_{\text{RLNC}}$.

For RLNC, recall that the minimum of coded transmissions, $U_{\text{RLNC}}$, is equal to the maximum number of lost packets across receivers in the systematic transmission phase, $W_{\max}$. The relationship between the mean of

$U_{\text{RLNC}}$ and $N$ can be written as:

$$\Pr(U_{\text{RLNC}} \leqslant W) = \Pr(W_{\max} \leqslant W) = \left( \sum_{i=1}^{W} \binom{K_T}{i} P_e^i (1-P_e)^{K_T - i} \right)^N \tag{10a}$$

$$E[U_{\text{RLNC}}] = E[W_{\max}] = \sum_{W=1}^{K_T} W \left( \Pr(W_{\max} \leqslant W) - \Pr(W_{\max} \leqslant W-1) \right) \tag{10b}$$

where $\Pr(W_{\max} \leqslant W)$ is the cumulative distribution function (CDF) of $W_{\max}$ with $W \in [0, K_T]$. This equation indicates that $U_{\text{RLNC}}$ increases quickly with $N$ when $N$ is relatively smaller than $K_T$ and increases gradually with $N$ when $N$ is relatively greater than $K_T$, as demonstrated in Fig. 2, where $K_T = 15$ and $P_e = 0.2$.

For IDNC, we know that with increasing $N$, more original packets may be wanted after the systematic transmission phase and thus the size of the conflict matrix $\boldsymbol{C}$ will increase. Furthermore, every new receiver may incur new conflicts between packets and thus the number of zeros in $\boldsymbol{C}$ will decrease with $N$. Both of these two factors will increase $U_{\text{IDNC}}$. However, there is no analytical expression for the relationship between the mean of $U_{\text{IDNC}}$ and $N$. As an alternative, we resort to numerical simulations. We randomly generate $M$ samples of the SFM $\boldsymbol{A}$ according to the system parameters $K_T$, $P_e$ and $N$. Then we average the $U_{\text{IDNC}}$ across the simulated SFMs to obtain the average $U_{\text{IDNC}}$.

The numerical results with $K_T = 15$, $P_e = 0.2$, $M = 10^5$ and $N \in [1, 45]$ are plotted in Fig. 2. Similar simulation is applied to RLNC and the average $U_{\text{RLNC}}$ is also plotted in this figure. As expected, the simulated average matches very well the theoretical mean, derived in (10). Our first observation is that $U_{\text{IDNC}}$ increases quickly with $N$ when $N$ is smaller than $K_T$ and then increases almost linearly with $N$ when $N$ is greater than $K_T$. The second observation is that $U_{\text{IDNC}}$ is almost the same as $U_{\text{RLNC}}$ when $N$ is smaller than $K_T$, but is greater than $U_{\text{RLNC}}$ when $N$ is greater than $K_T$. This gap grows gradually with $N$.

Therefore, we conclude that IDNC is a good alternative to RLNC in terms of the absolute minimum number of coded transmissions when the number of receivers $N$ is smaller than $K_T$, but is inefficient when $N$ becomes larger.

## B. The Distribution of $V_{\text{IDNC}}$ and $V_{\text{RLNC}}$

### 1) The Distribution of $V_{\text{IDNC}}$:

For a given SFM $\boldsymbol{A}$, we first consider the probability that a receiver $R_n$ still needs $V_n$ original packets after the first round of coded transmissions. This is denoted by $\Pr(V_n)$, where $V_n$ ranges from 0 (no more transmission needed) to $W_n$ (receiver $R_n$ did not receive any innovative packets during $U_{\text{IDNC}}$ transmissions due to poor channel conditions).

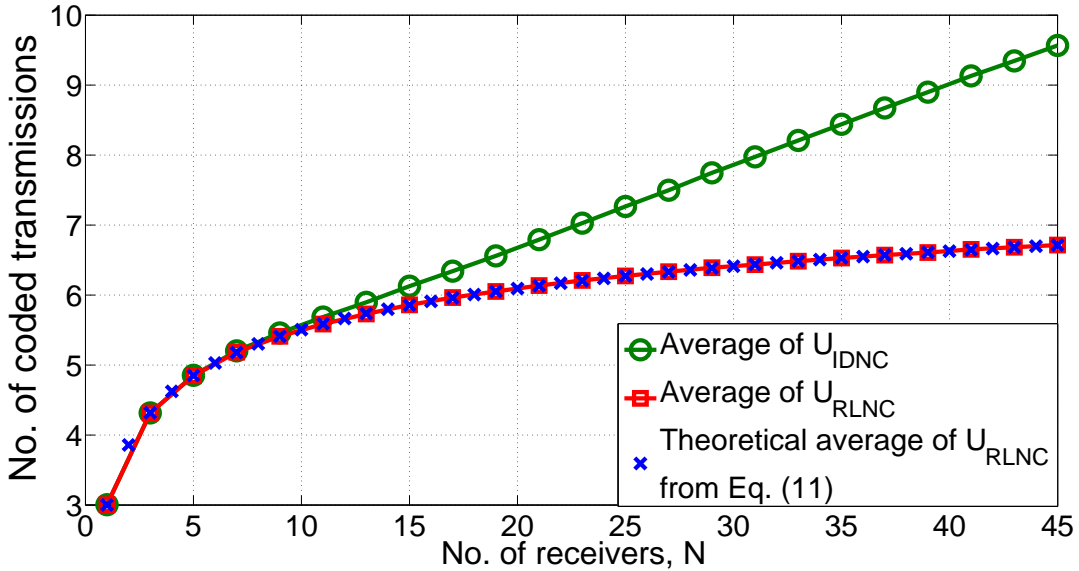

**Fig. 2:** Average $U_{\text{IDNC}}$ and $U_{\text{RLNC}}$ versus $N$ ($K_T = 15$, $P_e = 0.2$).

For obtaining $\Pr(V_n)$, we assume a specific minimal collection $\mathcal{S}$ or $\boldsymbol{A}$ with a size of $U_{\text{IDNC}}$ was chosen for the first round of coded transmission, where $d_k$ is the diversity of $P_k$ within this minimal collection. We denote the $n$-th row of the SFM $\boldsymbol{A}$ after the systematic phase by $\boldsymbol{a}_n$, which is the state feedback vector of $R_n$. The number of ones in $\boldsymbol{a}_n$ is represented by $\|\boldsymbol{a}_n\|_1$, where $\|\cdot\|_1$ means 1-norm and we have $\|\boldsymbol{a}_n\|_1 = W_n$. We further denote the $n$-th row of the updated $\boldsymbol{A}$ after the first round of coded transmission by $\boldsymbol{a}'_n$. It may take any form under the constraint that $\boldsymbol{a}'_n \leqslant \boldsymbol{a}_n$, i.e., every entry in $\boldsymbol{a}'_n$ is smaller than or equal to the corresponding entry in $\boldsymbol{a}_n$. Then, the general expression of $\Pr(V_n)$ can be written as:

$$\Pr(V_n = V) = \sum_{\forall \boldsymbol{a}'_n : \|\boldsymbol{a}'_n\|_1 = V} \left( \prod_{\forall k: P_k \in \mathcal{W}_n} \left((P_e)^{d_k}\right)^{a'_n(k)} \left(1 - (P_e)^{d_k}\right)^{1-a'_n(k)} \right) \quad (11)$$

where $a'_n(k)$ is the $k$-th entry of $\boldsymbol{a}'_n$.

For example, suppose that $K = 4$ after the systematic phase, and $R_n$ still wants $P_1$, $P_3$ and $P_4$, i.e., $\boldsymbol{a}_n = [1\ 0\ 1\ 1]$. Suppose that $R_n$ still wants $V_n = 2$ packets after the first round of coded transmission. Then there are three possibilities for $\boldsymbol{a}'_n$ as follows: $\boldsymbol{a}'_n = [0\ 0\ 1\ 1]$, $\boldsymbol{a}'_n = [1\ 0\ 0\ 1]$ and $\boldsymbol{a}'_n = [1\ 0\ 1\ 0]$. The probability of $\boldsymbol{a}'_n = [0\ 0\ 1\ 1]$ is the product of the probabilities of three independent events: $P_1$ has been received by $R_n$, but $P_3$ and $P_4$ have not. Their corresponding probabilities are $1 - P_e^{d_1}$, $P_e^{d_3}$ and $P_e^{d_4}$, respectively.[5] By summing up

---

[5] For example, for $P_1$ to be received by $R_n$ it suffices that the receiver did not experience erasures during at least one out of $d_k$ time slots where $P_1$ was broadcast. Also, since $a'_n(k) = 0$ or $a'_n(k) = 1$, the appropriate event probability in (11) is automatically chosen.

the probabilities of the three possible $\boldsymbol{a}'_n$, we obtain $\Pr(V_n = 2)$.

**Remark 4.** *From* (11)*, the significance and impact of packet diversity on IDNC overall throughput should become apparent. The larger the packet diversity, the higher the probability that a receiver successfully receives its requested packet during the first $U_{\text{IDNC}}$ coded transmissions and hence, the lower probability that more transmissions are required in the second round. This is the main reason why the sender should send maximal encoding sets (rather than disjoint sets) in a minimal collection with the largest average diversity across all receivers as suggested by* (9).

By employing (11) we are able to derive the distribution of $\Pr(V_{\text{IDNC}})$:

- Having computed $\Pr(V_n = 0)$ for a given $R_n$, the probability that no receiver needs more transmissions beyond $U_{\text{IDNC}}$, is simply

$$\Pr(V_{\text{IDNC}} = 0) = \prod_{n=1}^{N} \Pr(V_n = 0) \tag{12}$$

- Having computed $\Pr(V_n = 0)$ and $\Pr(V_n = 1)$, the cumulative distribution function (CDF) of $V_n$ for $R_n$ is:

$$\Pr(V_n \leqslant 1) = \Pr(V_n = 0) + \Pr(V_n = 1) \tag{13}$$

Now given $\Pr(V_n \leqslant 1)$ for all the receivers $R_n \in \mathcal{R}$, the CDF of $V_{\text{IDNC}}$ for the system is:

$$\Pr(V_{\text{IDNC}} \leqslant 1) = \prod_{n=1}^{N} \Pr(V_n \leqslant 1) \tag{14}$$

- Similarly, we can calculate the CDF value $\Pr(V_n \leqslant 2)$ for a given $R_n$. If every receiver needs at most two packets, the probability that the system needs at most two extra transmissions is *upper bounded by*

$$\Pr(V_{\text{IDNC}} \leqslant 2) \leqslant \prod_{n=1}^{N} \Pr(V_n \leqslant 2) \tag{15}$$

The inequality comes from the fact that if more than 2 receivers need two packets, we might need 3 or more transmissions to accommodate this using IDNC.

- Iteratively, we can compute $\Pr(V_n \leqslant 3)$ and so on up to $\Pr(V_n \leqslant W_n)$. Notice that $\Pr(V_n \leqslant W_n + 1) = \cdots = \Pr(V_n \leqslant U_{\text{IDNC}}) = 1$. Then an upper-bound on the CDF, $\Pr(V_{\text{IDNC}} \leqslant V)$, for any $V \in [3, U_{\text{IDNC}}]$ is obtained.

Consequently, we estimate the PDF of $V_{\text{IDNC}}$ as:

$$\Pr(V_{\text{IDNC}} = V) = \begin{cases} \Pr(V_{\text{IDNC}} \leqslant V), & V = 0 \\ \Pr(V_{\text{IDNC}} \leqslant V) - \Pr(V_{\text{IDNC}} \leqslant V - 1), & V \in [1, U_{\text{IDNC}}] \end{cases} \tag{16}$$

We note that $\Pr(V_{\text{IDNC}} = V)$ for $V \geqslant 3$ is an approximation rather than a provable bound. In order to verify its accuracy, we run some simulations as follows:

1) Randomly generate an SFM $\boldsymbol{A}$ and find out the minimal collection $\mathcal{S}$ of $\boldsymbol{A}$ which has the largest average diversity.

2) Transmit the maximal encoding sets in $\mathcal{S}$ through erasure links in the first round of coded transmission phase. Collect feedback and update $\boldsymbol{A}$, denoted by $\boldsymbol{A}'$.

3) Calculate the absolute minimum number of coded transmissions of $\boldsymbol{A}'$, the result is an instance of $V_{\text{IDNC}}$ of $\boldsymbol{A}$.

4) Repeat steps 2-3 $M$ times and obtain $M$ instances of $V_{\text{IDNC}}$.

5) Work out the numerical PDF of $V_{\text{IDNC}}$ using these $M$ instances.

The approximate PDF of $V_{\text{IDNC}}$ of this $\boldsymbol{A}$ is calculated using (16) and is compared with the numerical PDF of $V_{\text{IDNC}}$. The mean-squared-error (MSE) of the approximate PDF is then obtained. This simulation is repeated for $10^3$ randomly generated $\boldsymbol{A}$. The average MSE with system parameters $K_T = 15$, $N = 10$, $P_e = 0.2$ and $M = 10^5$ is $2.44 \times 10^{-4}$. Similar simulations on RLNC show an average MSE of $2.13 \times 10^{-4}$. Both approximation errors are acceptable because the theoretical PDF of $V_{\text{RLNC}}$ (which will be derived next) is exact. Therefore, the approximate PDF of $V_{\text{IDNC}}$ that we derived in (16) is reliable. Fig. 3 demonstrates the simulation results for a random sample of $\boldsymbol{A}$.

## C. The Distribution of $V_{\text{RLNC}}$

Similar to the case of IDNC, we also consider the probability that a receiver $R_n$ still needs $V_n$ transmissions after the first round comprising $U_{\text{RLNC}} = W_{\max}$ coded transmissions. Denote this probability by $\Pr(V_n)$, where $V_n$ ranges from 0 to $W_n$. Generally, it can be calculated as:

$$\Pr(V_n = V) = \begin{cases} \sum_{w=W_n}^{W_{\max}} \binom{W_{\max}}{w}(P_e)^{W_{\max}-w}(1-P_e)^w, & V = 0 \\ \binom{W_{\max}}{W_n-V}(P_e)^{W_{\max}-W_n+V}(1-P_e)^{W_n-V}, & V \in [1, W_n] \end{cases} \quad (17)$$

To demonstrate this equation, we consider the cases of $V_n = 0, 1, 2$. Then we employ them to derive the distribution of $V_{\text{RLNC}}$.

- $\Pr(V_n = 0)$ is the probability that $R_n$ has received *at least* $W_n$ encoded packets during $W_{\max}$ RLNC transmissions. Hence the first line of Eq. (17). Then the probability that no receiver needs more transmissions

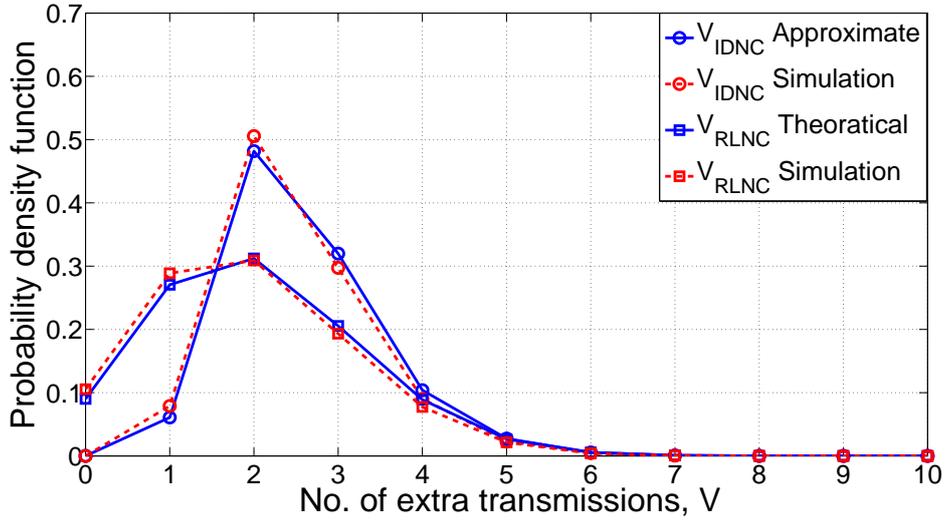

**Fig. 3:** Theoretical and numerical PDF of $V_{\text{IDNC}}$ and $V_{\text{RLNC}}$ for a randomly generated SFM $A$ ($K_T = 15, P_e = 0.2$).

after the first round of coded transmission is simply

$$\Pr(V_{\text{RLNC}} = 0) = \prod_{n=1}^{N} \Pr(V_n = 0) \tag{18}$$

- $\Pr(V_n = 1)$ is the probability that $R_n$ has received $W_n - 1$ encoded packets during $W_{\max}$ RLNC transmissions. Similar to the operations in the IDNC case, we can write:

$$\Pr(V_n = 1) = \binom{W_{\max}}{W_n - 1}(P_e)^{W_{\max}-(W_n-1)}(1-P_e)^{W_n-1} \tag{19}$$

$$\Pr(V_n \leqslant 1) = \Pr(V_n = 0) + \Pr(V_n = 1) \tag{20}$$

$$\Pr(V_{\text{RLNC}} \leqslant 1) = \prod_{n=1}^{N} \Pr(U_n \leqslant 1) \tag{21}$$

where $\Pr(V_{\text{RLNC}} \leqslant 1)$ is the cumulative probability that at most one more transmission beyond $U_{\text{RLNC}}$ is needed by the system.

- Iteratively, we can compute $\Pr(V_n \leqslant 2)$ and so on up to $\Pr(V_n \leqslant W_n)$. Notice that $\Pr(V_n \leqslant W_n + 1) = \cdots = \Pr(V_n \leqslant U_{\text{RLNC}}) = 1$. Then the CDF $\Pr(V_{\text{RLNC}} \leqslant V)$ for any $V \in [0, U_{\text{RLNC}}]$ can be calculated and there is no approximation involved for RLNC.

Consequently, the theoretical PDF of $V_{\text{RLNC}}$ can be calculated through subtracting adjacent CDF values of $V_{\text{RLNC}}$ in the same way as in (16). This PDF is exact. An example of the theoretical and numerical PDF of $V_{\text{RLNC}}$ are plotted in Fig. 3 and they well match each other.

*D. Numerical Comparison*

In this subsection, we combine the number of transmissions in the first and second rounds of coded transmission, which is denoted by $H_{\text{IDNC}}$ and $H_{\text{RLNC}}$ for IDNC and RLNC schemes, respectively. The PDF of $H$ reveals the relative throughput performance of IDNC and RLNC. For a given SFM $\boldsymbol{A}$, the PDF of $H_{\text{IDNC}}$ is expressed as:

$$\Pr(H_{\text{IDNC}} = U) = \begin{cases} 0, & U \in [0, U_{\text{IDNC}} - 1] \\ \Pr(V_{\text{IDNC}} = U - U_{\text{IDNC}}), & U \in [U_{\text{IDNC}}, 2U_{\text{IDNC}}] \end{cases} \quad (22)$$

Similarly, the PDF of $H_{\text{RLNC}}$ is:

$$\Pr(H_{\text{RLNC}} = U) = \begin{cases} 0, & U \in [0, U_{\text{RLNC}} - 1] \\ \Pr(V_{\text{RLNC}} = U - U_{\text{RLNC}}), & U \in [U_{\text{RLNC}}, 2U_{\text{RLNC}}] \end{cases} \quad (23)$$

We then numerically obtain the average PDF of both $H_{\text{IDNC}}$ and $H_{\text{RLNC}}$. The results for a system with $K_T = 15$, $P_e = 0.2$ and $N = 5, 15, 20, 30$ are plotted in Fig. 4. When $N$ is as small as 5, the throughput performance of IDNC is almost the same as that of RLNC. When $N$ is equal to $K_T$, there is only marginal throughput performance gain of RLNC over IDNC. When $N$ further increases, the gap between them grows. The $H_{\text{IDNC}}$ having the largest PDF value under $N = 30$ is 9, while the $H_{\text{RLNC}}$ having the largest PDF value under $N = 30$ is 7, indicating that IDNC in general may require 2 more coded transmissions than RLNC after the systematic phase. Given the better decoding delay performance of IDNC to be examined in the next section, we conclude that IDNC is a good alternative to RLNC when $N$ is relatively smaller than $K_T$. Recall that the gap between $U_{\text{IDNC}}$ and $U_{\text{RLNC}}$ under $N = 30$ is 1.5, as shown in Fig. 2. We further conclude that the gap between $U_{\text{IDNC}}$ and $U_{\text{RLNC}}$ is the dominant factor on the gap between $H_{\text{IDNC}}$ and $H_{\text{RLNC}}$, while the gap between $V_{\text{IDNC}}$ and $V_{\text{RLNC}}$ plays a secondary role.

## VI. The Advantage of IDNC over RLNC: Packet Decoding Delay Analysis

In the previous section, by showing that RLNC in general requires less block completion delay than IDNC, we confirmed that the throughput performance of IDNC never exceeds RLNC. In this section, we consider another type of delay, called packet decoding delay, for which IDNC may be better than RLNC. It is defined as follows:

**Definition 14.** *Packet decoding delay is the number of transmissions after which a wanted original data packet can be decoded by a receiver.*

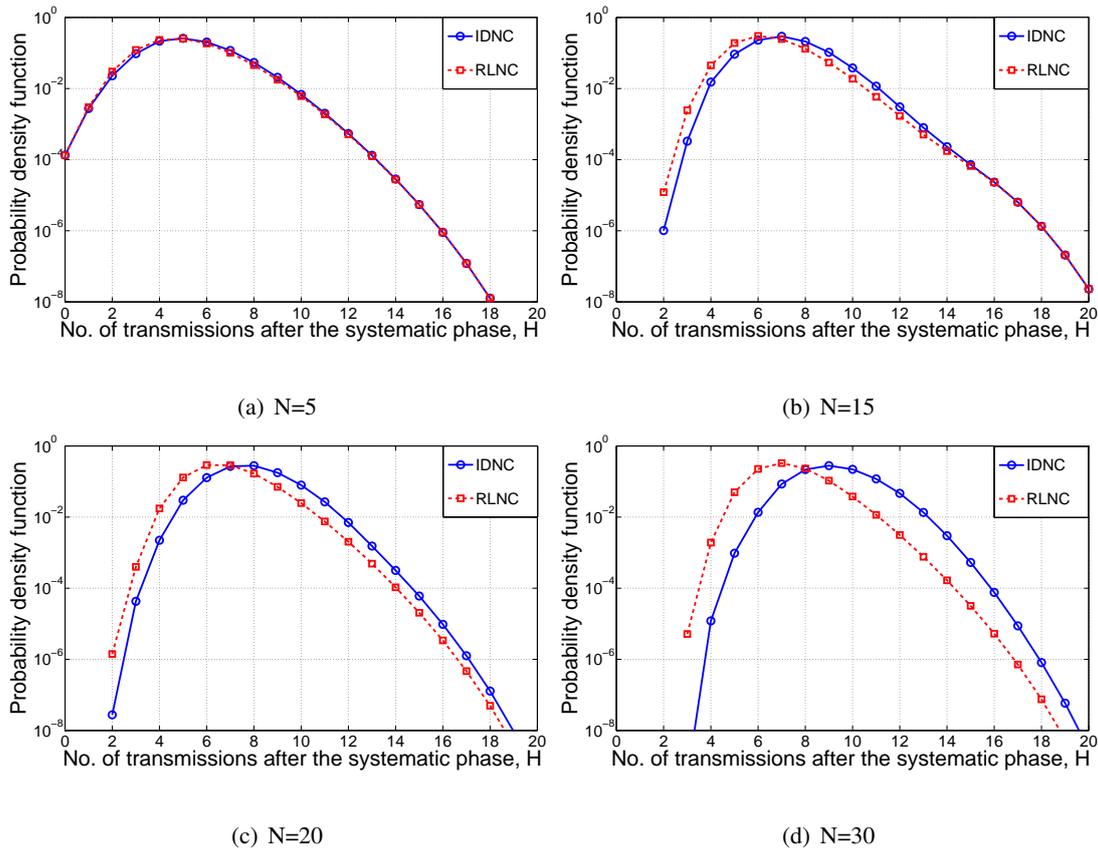

**Fig. 4:** Throughput comparison between IDNC and RLNC with $K_T = 15$, $P_e = 0.2$ and different number of receivers $N$.

This type of delay metric is important for determining IDNC and RLNC performance for intermediate packet delivery and especially in applications where packet delivery can occur regardless of packet order, such as multiple-description source coding [9], [11], [17].

Because IDNC and RLNC share the systematic transmission phase, we only consider original data packets which are still wanted after the systematic transmission phase, i.e. packets in $\mathcal{P}_K$. Furthermore, which packets will be still wanted after the first round of coded transmission phase depend on the channel erasure patterns during this round, which makes the decoding delay analysis intractable. Therefore, we only consider the first round of coded transmission phase and thus study the packet decoding delays in the closed interval $[1, U_{\text{IDNC}}]$ for IDNC and $[1, U_{\text{RLNC}}]$ for RLNC.

We then define a type of conceptual packet called *algorithmic packet* to enable the calculation of packet decoding delay. Its intuition comes from the fact that if an original packet has $T$ targeted receivers, its decoding delay at these $T$ receivers should be counted separately.

**Definition 15.** *Algorithmic packets corresponding to an original data packet, $P_k$, are the duplicates of that packet at its targeted receivers. The receiving probabilities of these packet duplicates are independent of each other.*

Given an SFM $\mathbf{A}$, there are $K$ original packets with $T_1, \cdots, T_K$ targeted receivers. Thus there are in total $T_{\text{total}} = \sum_{k=1}^{K} T_k$ algorithmic packets. They are denoted by $\mathcal{Q} = \{Q_1, Q_2, \cdots, Q_{T_{\text{total}}}\}$, where $Q_1, \cdots, Q_{T_1}$ are duplicates of $P_1$ at the receivers in $\mathcal{T}_1$, $Q_{T_1+1}, \cdots, Q_{T_1+T_2}$ are duplicates of $P_2$ at the receivers in $\mathcal{T}_2$, and so on. Their decoding delays are denoted by $l_1, \cdots, l_{T_{\text{total}}}$, respectively, whose values can range from 1 to $U_{\text{IDNC}}$ for IDNC and from 1 to $U_{\text{RLNC}}$ for RLNC, respectively. In this section we aim to propose useful metrics to measure the decoding delay of these algorithmic packets and investigate how IDNC could outperform RLNC in terms of these metrics. As a starting point, we calculate the *average packet decoding delay*:

$$L = \frac{\sum_{k=1}^{T_{\text{total}}} l_k}{T_{\text{total}}} \tag{24}$$

Denote the number of decoded algorithm packets in the $u$-th transmission by $D_u$, where $u \leqslant U_{\text{IDNC}}$ or $u \leqslant U_{\text{RLNC}}$. Then, the average packet decoding delay can also be calculated as:

$$L = \frac{\sum_{u=1}^{U} u D_u}{T_{\text{total}}} \tag{25}$$

where $U = U_{\text{IDNC}}$ or $U_{\text{RLNC}}$. In the best case scenario where channel links are erasure-free, average packet decoding delay of both IDNC and RLNC can be easily calculated and compared given the SFM $\mathbf{A}$. Below is an example:

**Example 4.**

*Consider the SFM $\mathbf{A}$ in Example 3 and its minimal collection $\mathcal{S} = \{\{P_1, P_3\}, \{P_4, P_6\}, \{P_2, P_3, P_5\}\}$. There are $T_{\text{total}} = 13$ algorithmic packets. We then demonstrate the decoding progress of IDNC and RLNC in the first round of coded transmission phase, respectively, assuming erasure-free packet reception at all receivers:*

- *IDNC: The first transmission will provide $T_1 + T_3 = 5$ algorithmic packets. The second transmission will provide $T_4 + T_6 = 5$ algorithmic packets. The final transmission will provide $T_2 + T_5 = 3$ algorithmic packets (notice that $P_3$ has already been decoded in the first transmission). The average packet decoding delay of this $\mathbf{A}$ if IDNC is employed will be:*

$$L_{\text{IDNC}} = \frac{5 \times 1 + 5 \times 2 + 3 \times 3}{13} \approx 1.85$$

*which means each algorithmic packet needs to wait about 1.85 transmissions for decoding.*

- *RLNC: In the first transmission, no receiver can decode anything because there is no receiver who wants only one packet. In the second transmission, both $R_2$ and $R_5$ can decode 2 algorithmic packets, in total 4* *algorithmic packets. In the final transmission, $R_1$, $R_3$ and $R_4$ can decode 3 algorithmic packets respectively, in total 9 algorithmic packets. The average packet decoding delay of this $\boldsymbol{A}$ if RLNC is employed will be:*

$$L_{\text{RLNC}} = \frac{0 \times 1 + 4 \times 2 + 9 \times 3}{13} \approx 2.7$$

*which means each algorithmic packet needs to wait about 2.7 transmissions for decoding. It is $46\%$ higher than $L_{\text{IDNC}}$.*

In the presence of link erasures, computing the exact value of $D_u$ is intractable. Instead, we determine the statistical *mean* of $D_u$, denoted by $E[D_u]$. Then the mean of average packet decoding delay is defined as:

$$E[L] = \frac{\sum_{u=1}^{U} u E[D_u]}{\sum_{u=1}^{U} E[D_u]} \quad (26)$$

where $U = U_{\text{IDNC}}$ or $U_{\text{RLNC}}$. We derive $E[D_u]$ for both IDNC and RLNC in the next subsection.

## A. $E[D_u]$ of IDNC and RLNC

### 1) $E[D_u]$ of IDNC:

Consider an SFM $\boldsymbol{A}$ and a given minimal collection $\mathcal{S} = \{\mathcal{M}_1, \mathcal{M}_2, \cdots, \mathcal{M}_{U_{\text{IDNC}}}\}$ (for example, the one that has the largest average diversity according to (9) and has its maximal encoding sets reordered to benefit most receivers first). In the $u$-th transmission, packets in $\mathcal{M}_u$ are encoded using $\mathbb{F}_2$ and transmitted. For each packet $P_{k'} \in \mathcal{M}_u$, its number of targeted receivers is denoted by $T_{k'}$ and its diversity within the group of *already* transmitted maximal encoding sets $\{\mathcal{M}_1, \cdots, \mathcal{M}_u\}$ is denoted by $d_{k'}(u)$. Consequently, each packet $P_{k'}$ provides $T_{k'}$ algorithmic packets, sharing independent and identical probability of being decoded by its targeted receiver:

$$\Pr(P_{k'} \text{ decoded at } u) = P_e^{d_{k'}(u)-1}(1 - P_e) \quad (27)$$

which means that in the first $d_{k'}(u) - 1$ transmissions about $P_{k'}$, a targeted receiver always experienced erasures. But in the $u$-th transmission, this receiver does not experience erasure. Consequently, the mean number of decoded algorithmic packets, $E[D_u]$, is:

$$E[D_u] = \sum_{k': P_{k'} \in \mathcal{M}_u} T_{k'} \Pr(P_{k'} \text{ decoded at } u) \quad (28)$$

**Remark 5.** *From (27) and (28), the significance of packet diversity and early transmission of packets requested by more receivers on the IDNC decoding delay performance should become apparent. The larger the packet diversity $d_{k'}(u)$, the higher the probability that a receiver has already successfully decoded its requested packet* *by time slot $u$. This generally enhances IDNC decoding delay performance and is the main reason why the sender should send maximal encoding sets (rather than disjoint sets) in a minimal collection with the largest average diversity across all receivers as suggested by* (9) *and should also send maximal encoding sets that target most receivers first.*

*2) $E[D_u]$ of RLNC:*

For a receiver $R_n \in \mathcal{R}$ who wants $W_n \leqslant u$ original data packets, the probability that it can decode $W_n$ packets in the $u$-th transmission is equal to the probability that it did not experience erasures in $W_n - 1$ out of the first $u - 1$ transmissions and is also not in erasure in the $u$-th transmission:

$$\Pr(W_n, u) = \binom{u-1}{W_n - 1} P_e^{u-W_n} (1 - P_e)^{W_n} \tag{29}$$

If $R_n$ can decode in the $u$-th transmission, it will generate $W_n$ decoded algorithmic packets. Thus the mean number of decoded packets that $R_n$ generates in the $u$-th transmission is $W_n \Pr(W_n, u)$. Consequently, the mean of $D_u$ across all receivers is:

$$E[D_u] = \sum_{R_n : W_n \leqslant u} W_n \Pr(W_n, u) \tag{30}$$

To this end, we have derived the $E[D_u]$ of a given SFM $\boldsymbol{A}$ under both IDNC and RLNC schemes. By substituting them into (26) we obtain the average packet decoding delay $E[L_{\text{IDNC}}]$ and $E[L_{\text{RLNC}}]$ of this $\boldsymbol{A}$. This metric could be a useful criterion for adaptively choosing between IDNC or RLNC after the systematic network coding phase.

**Remark 6.** *Although we have only computed the mean decoding delay across receivers,* (27) *and* (29) *completely characterize probabilistic decoding delay performance of IDNC and RLNC. For example, it is possible to compute higher order statistics of the decoding delay.*

*B. The impact of number of receivers on the average packet decoding delay*

In this subsection, we investigate how the number of receivers, $N$, impacts the decoding delay performance of IDNC and RLNC schemes. Because the derivation of both $E[L_{\text{IDNC}}]$ and $E[L_{\text{IDNC}}]$ is based on a specific $\boldsymbol{A}$, the investigation is partly through simulations. For a given set of system parameters $K_T$, $P_e$ and $N$, $M$ samples of $\boldsymbol{A}$ are randomly generated and their $E[L_{\text{IDNC}}]$ and $E[L_{\text{RLNC}}]$ are then analytically calculated. The average

4of the $M$ values of $E[L_{\text{IDNC}}]$ is the average $E[L_{\text{IDNC}}]$ under the current set of system parameters and so is the average $E[L_{\text{RLNC}}]$.

Fig. 5 shows the numerical results of the average $E[L_{\text{IDNC}}]$ and $E[L_{\text{RLNC}}]$ with parameters $K_T = 15$, $P_e = 0.2$, $M = 10^5$ and $N$ ranging from 1 to 45 with a step of 2. Our first observation is that the decoding delay profile of IDNC is similar to the throughput delay profile of IDNC. That is, $E[L_{\text{IDNC}}]$ increases almost linearly with $N$, from 2.7 when $N = 5$ to 4.7 when $N = 45$. Meanwhile, the average packet decoding delay of RLNC increases slowly with $N$ and the increment is marginal when $N$ is greater than $K_T$. The second observation is that, when $N$ is smaller than $K_T$, the average decoding delay of RLNC is between 30% to 50% greater than IDNC. It indicates a significant decoding delay performance gain of IDNC over RLNC. This gain decreases gradually with $N$ and both IDNC and RLNC have a average decoding delay of 4.2 when $N = 33$, which is just above twice of $K_T$. After that, the decoding delay performance of RLNC is better than IDNC. Interestingly, we can attribute the better performance of RLNC over IDNC in the last regime where $N$ is significantly larger than $K_T$ to its better throughput (block completion delay) performance as observed in Fig. 2. In other words, when $N >> K_T$, $U_{\text{IDNC}} > U_{\text{RLNC}}$, which postpones the decoding opportunity of some packets to much further time slots in IDNC. This clearly shows the interplay between throughput and decoding delay in network coded schemes, which is often very hard to characterize.

Recall that IDNC is a good alternative to RLNC in terms of throughput when $N$ is relatively smaller than $K_T$. Considering both the throughput and decoding delay performance, we make the following three conclusions from a statistical perspective: 1) When $N$ is relatively smaller than $K_T$, IDNC is more preferable than RLNC; 2) When $N$ is roughly in the range $[K_T, 2K_T]$, IDNC and RLNC lead in decoding delay and throughput, respectively, which scheme to choose depends on the specific application; and 3) When $N$ is larger than $2K_T$, RLNC is a better choice.

## VII. Conclusion

In this paper, we provided a systematic framework for throughput and decoding delay analysis and comparison of random linear network coding (RLNC) and instantly decodable network coding (IDNC) schemes in erasure-prone packet broadcast scenarios. Our approach was based on analyzing the two schemes under the same packet reception status after the systematic transmission phase. For IDNC, which was the harder problem to solve, this was enabled through a comprehensive graph theoretical study of the IDNC model and its encoding process. In particular, we presented easily-calculable upper bound and lower bounds on the minimum number of coded transmissions, $U_{\text{IDNC}}$ based on the packet conflict matrix. We also presented an optimum IDNC encoding algorithm to minimize $U_{\text{IDNC}}$. We introduced the notion of packet diversity, which we showed to be a key factor in determining the probability of needing residual coded transmissions beyond $U_{\text{IDNC}}$ and its decoding delay performance. Similar throughput and decoding delay metrics were derived (with more ease) for the RLNC counterpart. Finally, we provided insights into the role of feedback frequency on the complexity and overall throughput performance of IDNC.

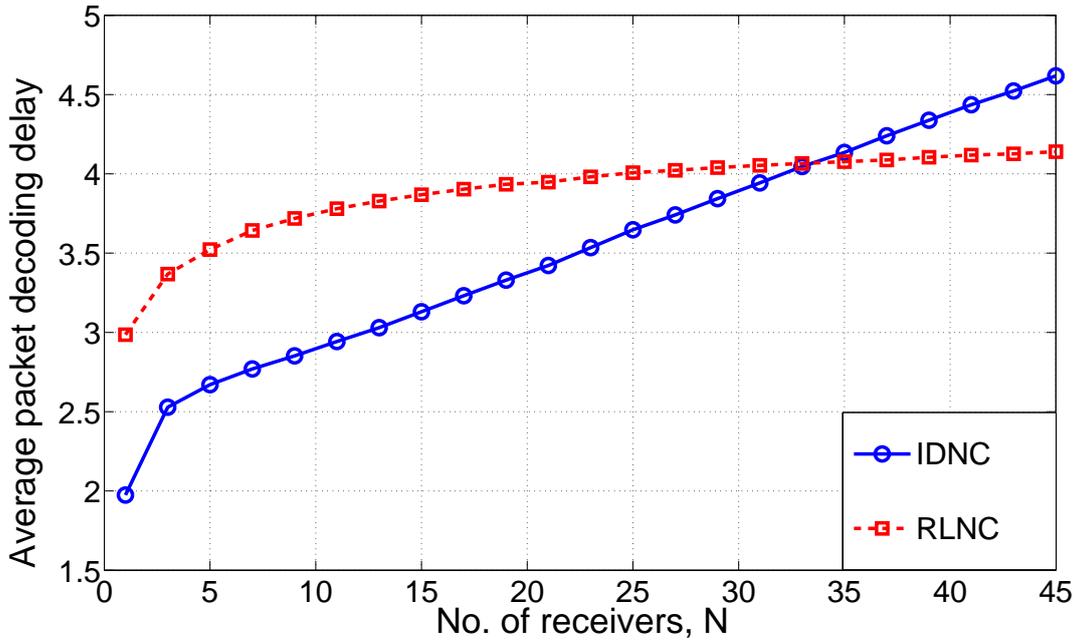

**Fig. 5:** Average packet decoding delay of IDNC and RLNC versus $N$ ($K_T = 15$, $P_e = 0.2$).

This comparative framework is useful for selecting between IDNC and RLNC given a specific instance of the packet reception status (SFM $\boldsymbol{A}$) depending on the specifically obtained throughput and delay performance metrics for each scheme. Furthermore, it enables statistical study of the impact of number of receivers on the average throughput and decoding delay. For example, we observed that both the throughput and decoding delay performance of IDNC degrades gradually as the number of receivers increases, while such degradation is much weaker in RLNC. Furthermore, IDNC is more preferable than RLNC when the number of receivers is smaller than the size of packet block, and the case reverses when the number of receivers is much greater than the size of packet block. In the middle regime, there is no clear winner as the throughput of RLNC is higher while the decoding delay of IDNC is lower.

# APPENDIX A

## PROOF OF THEOREM 2

We prove that if there is a collection $\mathcal{S} = \{\mathcal{M}_1, \cdots, \mathcal{M}_U\}$ satisfying the diversity constraint, there is also a $U$-coloring solution of $\overline{\mathcal{G}}$ and *vice versa*. An eligible $U$-coloring solution is denoted by $\{\mathcal{K}_1, \cdots, \mathcal{K}_U\}$ and has the following three properties: 1) the vertices in the same set $\mathcal{K}_i$ share the same color; 2) any two of vertices in the same $\mathcal{K}_i$ are not connected; 3) the intersection between any two sets $\mathcal{K}_i$ and $\mathcal{K}_j$ is empty.

If $\mathcal{S} = \{\mathcal{M}_1, \cdots, \mathcal{M}_U\}$ satisfies the diversity constraint, we can always construct a new group of vertex sets as follows, which also satisfy the diversity constraint: Set $\mathcal{K}_1 = \mathcal{M}_1$, then sequentially $\mathcal{K}_2 = \mathcal{M}_2 \setminus \mathcal{K}_1$, $\mathcal{K}_3 = \mathcal{M}_3 \setminus \bigcup\{\mathcal{K}_1, \mathcal{K}_2\}, \cdots, \mathcal{K}_U = \mathcal{M}_U \setminus \bigcup_{i=1}^{U-1}\{\mathcal{K}_i\}$. Following this construction process, it is clear that the intersection between any two sets $\mathcal{K}_i$ and $\mathcal{K}_j$ is empty, thus property 3) is satisfied. Because $\mathcal{K}_i \subseteq \mathcal{M}_i$, every vertex in the same $\mathcal{K}_i$ is connected to each other under $\mathcal{G}$. Therefore, they are all disconnected under $\overline{\mathcal{G}}$ and property 2) is satisfied. Hence, if we assign $U$ colors to $K_i, \cdots, \mathcal{K}_U$, they form a $U$-coloring solution of $\overline{\mathcal{G}}$.

Conversely, if $\{\mathcal{K}_1, \cdots, \mathcal{K}_U\}$ is a $U$-coloring solution of $\overline{\mathcal{G}}$, it is also a minimum clique cover solution of $\mathcal{G}$. For each $\mathcal{K}_i$, we can find a maximal clique $\mathcal{M}_i$ where $\mathcal{K}_i \subseteq \mathcal{M}_i$, then $\mathcal{S} = \{\mathcal{M}_1, \cdots, \mathcal{M}_U\}$ satisfies the diversity constraint.

Hence, the minimum collection size of the conflict matrix $\boldsymbol{C}$ equals the chromatic number of $\overline{\mathcal{G}}$.

# APPENDIX B

## PROOF OF THEOREM 3

Suppose $\mathcal{M}$ is a maximal clique of $\mathcal{G}$ and the chromatic number of $\overline{\mathcal{G}}$ is $U$. Assuming the chromatic number of $\overline{\mathcal{G}'} = \overline{\mathcal{G}} \setminus \mathcal{M}$ is $U' < U - 1$, there exists at least one $U'$-coloring solution of $\overline{\mathcal{G}'}$, denoted by $\{\mathcal{K}_1, \cdots, \mathcal{K}_{U'}\}$. If this is the case, $\{\mathcal{K}_1, \cdots, K_{U'}, \mathcal{M}\}$ is a eligible $(U'+1)$-coloring of $\overline{\mathcal{G}}$ and $U' + 1 < U$, contradicting the assumption that the chromatic number of $\overline{\mathcal{G}}$ is $U$. Thus the chromatic number of $\overline{\mathcal{G}'}$ is at least $U - 1$.

The chromatic number of $\overline{\mathcal{G}'}$ will still be $U$ if the removed $\mathcal{M}$ does not belong to any minimal collection of $\boldsymbol{C}$. This is equivalent to saying that there is no $\mathcal{K} \subseteq \mathcal{M}$ in any $U$-coloring solution of $\overline{\mathcal{G}}$, because otherwise we could obtain a minimal collection $\mathcal{S}$ of $\boldsymbol{C}$ comprising $\mathcal{M}$ using the $U$-coloring solution, which includes this $\mathcal{K}$ in the way we described in Appendix A. Hence, there are at least two colors involved in the vertices of $\mathcal{M}$ and there are vertices outside $\mathcal{M}$ with the same colors. Hence, removing vertices in $\mathcal{M}$ from $\overline{\mathcal{G}}$ cannot reduce the chromatic number of the resulted $\overline{\mathcal{G}'}$ to less than $U$.

## APPENDIX C

## AN EXAMPLE FOR $U_{\text{IDNC}}^{\text{online}} < U_{\text{IDNC}}^{\text{semi-online}}$

Consider an initial SFM $\boldsymbol{A}$ with three wanted packets and three receivers:

$$\boldsymbol{A} = \begin{array}{c|ccc|c} & P_1 & P_2 & P_3 & \\ \hline & 1 & 1 & 0 & R_1 \\ & 1 & 0 & 1 & R_2 \\ & 0 & 1 & 1 & R_3 \end{array}$$

and a channel erasure pattern with 5 future transmissions:

| $u$ | 1 | 2 | 3 | 4 | 5 |
|---|---|---|---|---|---|
| $R_1$ | X | O | O | O | O |
| $R_2$ | X | O | X | O | O |
| $R_3$ | O | O | O | O | O |

where $u$ is the transmission time slot index, $X$ means erasure and $O$ means successful packet reception.

We then describe the transmission progress in both semi-online feedback and fully-online feedback cases:

- Semi-online feedback

1) $P_1, P_2$ and $P_3$ all conflict with each other. Thus $\mathcal{S} = \{\{P_1\}, \{P_2\}, \{P_3\}\}$ and $U_{\text{IDNC}}^1 = 3$ in the first coded transmission round. $P_1$ is transmitted first, followed by $P_2$ and $P_3$.

2) After 3 transmissions, the sender collects feedback and the updated $\boldsymbol{A}'$ is:

$$\boldsymbol{A}' = \begin{array}{c|ccc|c} & P_1 & P_2 & P_3 & \\ \hline & 1 & 0 & 0 & R_1 \\ & 1 & 0 & 1 & R_2 \\ & 0 & 0 & 0 & R_3 \end{array}$$

We note that $W_2 = 2$ and it can be easily verified that $U_{\text{IDNC}}^2 = 2$ in the second coded transmission round. Then $P_1$ is transmitted again in the fourth transmission, followed by $P_3$. After that, all the three receivers' demands are satisfied.

Thus the semi-online feedback scheme requires $U_{\text{IDNC}}^{\text{semi-online}} = U_{\text{IDNC}}^1 + U_{\text{IDNC}}^2 = 5$ coded transmissions.

- Fully-online feedback

  1) After transmitting $P_1$ in the first transmission, the sender collects feedback and the updated $\boldsymbol{A}'$ is equal to $\boldsymbol{A}$;

  2) Thus in the second transmission, $P_1$ will be transmitted again. The sender then collects feedback and the updated $\boldsymbol{A}'$ is:

  $$\boldsymbol{A}' = \begin{array}{c} \\ \\ \\ \\ \end{array} \begin{array}{ccc|c} P_1 & P_2 & P_3 & \\ \hline 0 & 1 & 0 & R_1 \\ 0 & 0 & 1 & R_2 \\ 0 & 1 & 1 & R_3 \end{array}$$

  3) In the third transmission $P_2$ is transmitted because $\mathcal{S}' = \{\{P_2\}, \{P_3\}\}$ for $\boldsymbol{A}'$. $P_2$ will be received by both $R_1$ and $R_3$ in this transmission and thus the only wanted packet after that is $P_3$.

  4) In the fourth transmission, $P_3$ is transmitted and will be successfully received by both $R_2$ and $R_3$.

Thus fully-online feedback scheme requires $U_{\text{IDNC}}^{\text{online}} = 4$ coded transmissions, which is smaller than $U_{\text{IDNC}}^{\text{semi-online}}$.